\documentclass[12pt]{iopart}

\usepackage{graphicx}
\usepackage{amsbsy}
\usepackage{color}

\newcommand{\eV}{\,\mathrm{eV}}
\newcommand{\GeV}{\,\mathrm{GeV}}

\newcommand{\cm}{\,\mathrm{cm}}

\newcommand{\yr}{\,\mathrm{yr}}
\newcommand{\Mpc}{\,\mathrm{Mpc}}
\newcommand{\kpc}{\,\mathrm{kpc}}
\newcommand{\pc}{\,\mathrm{pc}}
\newcommand{\muG}{\,\mu\mathrm{G}}

\newcommand{\tu}[1]{\mathrm{#1}}

\begin{document}
\title{Numerical propagation of high energy cosmic rays in the Galaxy
  I: technical issues}

\author{Daniel De Marco\dag, Pasquale Blasi\ddag\ and Todor Stanev\dag}

\address{\dag\ Bartol Research Institute, University of Delaware
Newark, DE 19716, U.S.A.}

\address{\ddag\ INAF/Osservatorio Astrofisico di Arcetri,
Largo E. Fermi, 5 - 50125 Firenze, ITALY}

\eads{\mailto{ddm@bartol.udel.edu}, \mailto{blasi@arcetri.astro.it},
\mailto{stanev@bartol.udel.edu}}

\begin{abstract}
We present the results of a numerical simulation of propagation of
cosmic rays with energy above $10^{15}$ eV in a complex magnetic
field, made in general of a large scale component and a turbulent
component. Several configurations are investigated that may represent
specific aspects of a realistic magnetic field of the Galaxy, though
the main purpose of this investigation is not to achieve a realistic
description of the propagation in the Galaxy, but rather to assess the
role of several effects that
 define the complex problem of
propagation. Our simulations of Cosmic Rays in the Galaxy will be
presented in Paper II. We identified several effects that
are difficult to interpret
 in a purely diffusive approach and that play a crucial
role in the propagation of cosmic rays in the complex magnetic field
of the Galaxy. We discuss at
length the problem of the extrapolation of our results to much lower
energies where data are available on the confinement time of cosmic
rays in the Galaxy. The confinement time and its dependence on
particles' rigidity are crucial ingredients for 1) relating the source 
spectrum to the observed cosmic ray spectrum; 2) quantifying the
production of light elements by spallation; 3) predicting the
anisotropy as a function of energy.
\end{abstract}

\maketitle

\section{Introduction}

A complete understanding of the origin of Cosmic Rays (CRs) will be
achieved when the acceleration processes, the sources and the
propagation from the sources to the Earth
will be included in a
self-consistent theoretical framework. This goal is far from being
achieved: for ultra high energy cosmic rays (UHECRs) the issue of the
propagation is probably easier to understand since the effect of
extragalactic magnetic field is expected to be not crucial, at least 
above energies of $\sim 4\times 10^{19}\eV$ \cite{BD04}. On the other hand in this
case the sources are fully unknown. For CRs below $\sim 10^7-10^8\GeV$
we are confident that the sources are located within our Galaxy and
most likely are supernova remnants (SNRs) \cite{galcr}. In this respect a large
bulk of information is being collected from X-ray and $\gamma$-ray
astronomy: high resolution X-ray observations have shown the presence
of intense magnetic fields in the vicinity of the shocks that bound
the shell of the remnant \cite{vbk}, thereby making the acceleration process
 easier to understand. The combination with the
observed X-ray spectra and the outstanding detection of $10-100$ TeV
gamma rays from a few SNRs \cite{funk} make a rather strong case in favor of these
astrophysical sources being the accelerators of protons up to the knee
or slightly above it \cite{galcr}. Nuclei would then be accelerated to even higher
maximum energies, up to $\sim 10^{17}$ eV for iron nuclei. Although
the observational situation and the theoretical understanding are both
experiencing a substantial improvement as far as the sources (or at
least SNRs) are concerned, a realistic description of the propagation
of CRs in the interstellar medium (ISM) is still missing, despite the
very impressive amount of work carried out on the topic (see
\cite{moska} and references therein for a recent review).
Such work may be classified in two broad classes: analytical
approaches and simulations. 

Most analytical work is based on the solution of the
diffusion-convection equation from a distribution of sources in a
medium with given diffusion properties. We include in this class the
work that is based on a numerical solution of the transport equation
(e.g. GALPROP \cite{galprop} or the model presented in
\cite{diffudrift}). In the most general case, the equation 
has been solved with both parallel and perpendicular diffusion taken
into account. These approaches start from the premise that the magnetic
field of the Galaxy induces only a diffusive behaviour on CRs, namely
the turbulent field is the key ingredient. This component is provided
{\it a priori}, either in the form of pre-calculated diffusion
coefficients or in the form of turbulent spectra. It is worth
stressing that the spectrum of the turbulence responsible for particle
diffusion, the total power in turbulent modes  and the origin of such
turbulence are unknown. However, if one assumes that the spectrum is
known, then the diffusion coefficients could be calculated, at least
in principle, using quasi-linear theory and neglecting the geometry of
the large scale background magnetic field. 

The most common approach in the literature consists of using low energy
data on the ratio of secondary to primary nuclei in CRs as a function
of energy to infer the energy dependence of the propagation time,
which in turn leads to a rough knowledge of the energy dependence of
the diffusion coefficient. Such dependence is then adopted in the
solution of the transport equation. 

The shortcomings and advantages of using the diffusion equation to
describe the propagation of CRs are easy to identify: this approach 
allows one to achieve a basic understanding of some issues (for
instance the spectral steepening induced by the particle
propagation and leakage from the Galaxy). Moreover the approach can be
used without limitations in the dynamical range
(particles from very low to very high momenta can be included).
On the other hand, the diffusion coefficients are given quantities;
even when the diffusion coefficients (as functions of particle
momentum and spatial location) are calculated from first principles
(quasi-linear theory) they are often used 
in regimes where the initial assumptions do not necessarily hold.
In addition, a multitude of effects related to
spatial gradients of the large scale fields are hardly accounted for. 

The numerical simulation of the propagation of CRs in arbitrary
magnetic fields solves part of these problems, but is limited by
the constraints on the computational time.
Previous investigations using this technique concentrated on very high energy
cosmic rays ($\sim10^{18-19}\eV$), and on the deflections produced by their
passage into the Galactic Magnetic Field \cite{gmfsimu} or on their
anisotropy around $10^{18}\eV$ \cite{leeclay}.
Other attempts investigated lower energies, e.g. Ref.~\cite{pt} was able
to reach down to $10^{17}\eV$, and calculated the times of escape from the
galaxy as a function of energy. The results obtained, however, seem to be
inconsistent with measurements at low energy. Indeed, in
Ref.~\cite{pt}, the escape time at $10^{17}\eV$ is found to be of the
order of $10^{5}\yr$ with an energy dependence of $E^{-1}$, much steeper
than the one expected for example from a normal Kolmogorov turbulence.
The extrapolation of this value to $10^{9}\eV$ produces a value several
orders of magnitude larger than the measured one. 
The problem of the steepness of the escape time in the simulations seems
to be a general one: it is present also in our simulations and seems
to continue to lower energies as we discuss below.

In this paper we describe the numerical code that we recently
completed for the propagation of cosmic rays in arbitrary magnetic
fields (both in their large scale and turbulent components). The code
represents a substantial improvement on previous efforts
in the same direction in several ways: first, we succeeded in
propagating the particles down to energies of
 $10^{14}$ eV,
 lower by at least one/two orders of magnitude compared with previous
simulations. Second, the turbulence responsible for diffusive particle
motion can be taken as three dimensional or one-dimensional, and as
isotropic or anisotropic. The large scale field is also completely
arbitrary. 

We present the results of this simulation effort in two papers. In the 
present paper (Paper I) we discuss all technical aspects and apply the
approach to several toy models of the magnetic field of the Galaxy in
order to emphasize the role of the physical effects that is necessary
to include in order to understand the propagation of CRs. In a second
paper (Paper II) we will describe the 
results of the simulation for given configurations of Galactic
magnetic fields which are commonly assumed as realistic. 

The paper is organized as follows: In \S \ref{sec:simula} we describe
the technical aspects of the simulation, with special attention to the
generation of the turbulent magnetic field. In \S \ref{sec:basic} we
illustrate some basic concepts of diffusion in the context of
quasi-linear theory, which allows us to define what is the common lore
of cosmic ray propagation in the Galaxy in terms of diffusion and
drifts. In \S \ref{sec:diffu} we describe the numerical procedure
adopted to calculate the parallel and perpendicular diffusion
coefficients. Finally, in \S \ref{sec:toy} we describe the results of
our computations for several toy models of the large scale field of
the Galaxy. We present our conclusions in \S \ref{sec:discussion}.

\section{Description of the simulation}\label{sec:simula}

We propagate particles in a magnetic field,
$\boldsymbol{B}=\boldsymbol{\delta B}+\boldsymbol{B_0}$, that is the sum
in each point of a regular and a random component. Both of them can in
principle depend on the position. 
As detailed in Ref.~\cite{clp} there are basically two methods to
implement the turbulent field: 1) pre-computing the field on a grid
using Fast Fourier Transform (FFT) 
and 2) calculating the field in each point along the particle
trajectory as the superposition of plane waves \cite{GJ}.

In the FFT approach the field is pre-computed on a grid in real space
from its power spectrum in reciprocal space. We set up a three dimensional
grid with integer coordinates from $0$ to $N-1$. Each vertex on the grid
corresponds to a wave vector $\boldsymbol{k}$ with components given by
the coordinates of the vertex. If any one of the components of
$\boldsymbol{k}$, for example $k_x$, is larger than $N/2$, then we
substitute it with $-(N-k_x)$ in order to take into account negative
frequencies. For each $\boldsymbol{k}$ we construct an amplitude vector,
$\boldsymbol{B}_{\boldsymbol{k}}$, with a length proportional to the
square root of the power in the corresponding mode: $k^{-(\gamma+2)/2}$, 
a random direction in the plane orthogonal to $\boldsymbol{k}$ and a
random phase.
Choosing the amplitude proportional to $k^{-(\gamma+2)/2}$ makes sure
that the power spectrum of the turbulent field is ${\cal P}(k)\propto
k^{-\gamma}$, whereas choosing the direction in the plane orthogonal to
$\boldsymbol{k}$ assures that $\boldsymbol{\nabla}\cdot
\boldsymbol{\delta B}=0$. We also have to make sure that 
$\boldsymbol{B}_{\boldsymbol{k}}$ satisfies the following condition
 for the resulting magnetic field to be real:
$\boldsymbol{B}_{(k_1,k_2,k_3)}=\boldsymbol{B}_{(N-k_1,N-k_2,N-k_3)}^*$.
The normalization is obtained by requiring that $\langle \delta B^2
\rangle = \sum B_{\boldsymbol{k}}^2$ and $B_{\boldsymbol{k}=(0,0,0)}$ is set to $0$ to
have $\langle \delta \boldsymbol{B} \rangle =0$.
At this point we compute the FFT \cite{fftw} and obtain the turbulent
field defined on a cubic grid with side $L_\tu{max}$ and spacing
$L_\tu{min}=L_\tu{max}/N$. We typically use $N=256$.

We assume the box is replicated periodically all over the simulation
volume and in order to calculate the turbulent field in a given point
the code uses the field value of the closest vertex. Another possibility
is to do an interpolation of the values at the eight vertexes surrounding
the point. We verified that the results obtained with the two methods
are equal on scales larger than the cell size ($L_\tu{min}$) and we
decided to use the former method.

The above description is valid for the general case of isotropic
turbulence. We also used 1D turbulence, a superposition of Alfven waves,
and in this case the generation proceeds along the same lines, but the
$\boldsymbol{k}$s are now chosen only parallel to the background field,
so that the fluctuating magnetic field is always perpendicular to it.
For the 1D field we typically use $N=4096$.

In the second approach the field is constructed as the sum of $N_m$
plane waves \cite{GJ,parizotdiffu}:
\begin{equation}\label{eq:gjdeltab}
  \boldsymbol{\delta B}=\sum_{n=1}^{N_m} A_{\boldsymbol{k}_n}
  \boldsymbol{\epsilon}_n \exp(\tu{i} k_n z'_n +\tu{i}\beta_n)\,,
\end{equation}
where $\boldsymbol{\epsilon}_n = \cos\alpha_n \boldsymbol{\hat{x}}'_n + \tu{i}
\sin\alpha_n \boldsymbol{\hat{y}}'_n$ and $\alpha_n$ and $\beta_n$ are
random phases. The primed coordinates are obtained by rotating the
reference frame so that the $z$ axis coincides with the direction of
propagation of the $n$-th wave, $\boldsymbol{k}_n$. The directions of
the $N_m$ waves are chosen randomly, while their amplitudes,
$A_{\boldsymbol{k}_n}$, are chosen as a function of $|\boldsymbol{k}_n|$  
according to the type of turbulence wanted. We follow Ref.~\cite{GJ} and
we use: 
\begin{equation}
  A^2_{\boldsymbol{k}}=\sigma^2 G( {\boldsymbol{k}} ) \big[ 
  \sum_{n=1}^{N_m} G(\boldsymbol{k}_n)
	\big]^{-1}\,,
\end{equation}
where
\begin{equation}
  G(\boldsymbol{k}) = \frac{\Delta V^{(d)}}{1+(k L_c)^{\gamma+(d-1)}}\,.	
\end{equation}
In these equations $\sigma$ fixes the normalization of the field,
$\sigma^2 = \langle \delta B^2 \rangle$, $L_c$ is the correlation
length, $\gamma$ is the slope of the turbulence power spectrum, $d$ is
its dimensionality and $\Delta V^{(d)}$ is the volume element for the
chosen dimensionality. In the present paper we use 3D and 1D turbulence
and in these cases $\Delta V^{(3)} = 4 \pi k^2 \Delta k$ and $\Delta
V^{(1)} = \Delta k$. The wavenumbers are chosen evenly spaced in
logarithmic scale between $k_\tu{min}$ and $k_\tu{max}$ and $\Delta k =
k \Delta \log k$.

The number of waves, $N_m$, used in the summation (\ref{eq:gjdeltab}),
is a key parameter and it should be large enough to reasonably describe
the turbulence. In Ref.~\cite{parizotdiffu} it was shown that if
$N_m$ is too small the transition from rectilinear to diffusive
propagation occurs on timescales much larger than the correct ones. It
was also found that a value of $100$ waves per decade is a reasonable
compromise between accuracy and computation time and we use this value
in our simulations. 

The only difference for the case of 1D turbulence is that instead of
choosing the $\boldsymbol{k}_n$ isotropically we choose them in the
direction parallel to the background field.

Both the methods described have their advantages and disadvantages: with
the FFT approach the time needed to obtain the turbulent field in a
given point is in general much smaller than in the plane wave approach.
In the first case all that is required is a lookup in a table (and
possibly some interpolations), whereas in the latter case there is a
summation over hundreds of waves to be done.
On the other hand, the dynamic range of the turbulence,
$L_\tu{max}/L_\tu{min}$, in the FFT approach is limited (at least in
the isotropic turbulence case) by the memory available to store the huge
matrices describing the magnetic field grid, whereas in the plane wave
approach the memory limitations are absent and the dynamic range can be
as big as required with the only limit given by the computation time.
As mentioned in Ref.~\cite{clp}, other limitations of the FFT approach
are inherent in its discreteness, $L_\tu{min}$, and in its limited size,
$L_\tu{max}$, and the results obtained with it can not be trusted when
the Larmor radius of the particles is smaller than $L_\tu{min}$ or
larger than $L_\tu{max}$.

\section{Basic facts about diffusion and drifts}\label{sec:basic}

In this section we summarize the basic facts on diffusion of cosmic
rays in a turbulent magnetic field superimposed to a large scale
spatially constant magnetic field $\boldsymbol{B_0}=B_0 \boldsymbol{\hat
z}$. Gradients in the large scale field induce drift motions of the
particles that add to the diffusive motion and in fact in some
circumstances may even dominate upon diffusion. 

\subsection{Diffusion}
\label{sec:diffu1}
In all the cases that we consider below we investigate 3D turbulence,
namely the perturbation of the large scale field has components both in
the plane perpendicular to $\boldsymbol{B_0}$ and along
$\boldsymbol{B_0}$. Therefore this case is somewhat more complex but
supposedly more realistic than the simpler case of Alfven waves
propagating along the
field $\boldsymbol{B_0}$ (we refer to this case as the {\it 1D case}),
typically considered in the literature on quasi-linear theory. 
In the case of 3D
turbulence, the perpendicular diffusion, though small compared with the
parallel diffusion in the quasi-linear regime, may become important for
the cases of strong turbulence $\delta B/B_0> 1$. On the basis of
quasi-linear theory the ratio of perpendicular to parallel diffusion
coefficient is given by
\begin{equation}
\frac{D_\perp}{D_\parallel} = \frac{1}{1+\left(
  \lambda_\parallel/r_L\right)^2}, 
\label{eq:ratio}
\end{equation} 
where the parallel pathlength is $\lambda_\parallel=3D_\parallel/v$,
$v$ is the particle velocity and $r_L$ is the Larmor radius in the
unperturbed magnetic field $B_0$. This expression remains valid as
long as $\delta B/B_0\ll 1$, but it also suggests that the
perpendicular diffusion coefficient approaches the parallel diffusion
coefficient in the regime of strong turbulence. In fact the real ratio
of the diffusion coefficients is affected by the random walk of the
field lines, which is not taken into account in Eq. \ref{eq:ratio}. 
This fact was found in \cite{JokipiiParker} and further discussed in
\cite{clp} 
and is illustrated in the next section in detail since it plays a
crucial role in the understanding of the results of the simulation of
cosmic rays in the Galaxy. 

The parallel diffusion coefficient can be estimated from the 1D case,
by using the quasi-linear theory:
\begin{equation}
D_\parallel = \frac{1}{3} r_L c \frac{1}{{\cal F}(k)},
\end{equation}
where ${\cal F}(k)=\left(\delta B (k)/B_0\right)^2$ is the normalized
power in the turbulent modes with wavenumber $k\propto 1/p$ resonant
with the particles with momentum $p$. Even in the 3D case this is a
reasonable approximation to the parallel diffusion coefficient since
this is dominated by the components of the perturbing field which are
perpendicular to the background field. In this case one can see that
the perpendicular diffusion coefficient is 
\begin{equation}
D_\perp \approx D_\parallel {\cal F}(k)^2.
\end{equation}
Since by definition ${\cal F}(k)\ll 1$ it is easy to see that
$D_\perp\ll D_\parallel$, which implies that in most cases the effect
of perpendicular diffusion is irrelevant if the propagation occurs
in the regime of weak turbulence. 

In numerical simulations of the propagation of cosmic rays in the
Galaxy it is usually assumed that $\delta B/B_0\sim {\cal O}(1)$.
This ratio is supported by general estimates, such as equipartition,
cosmic ray behavior and observations of total magnetic field in 
elliptical galaxies~\cite{Beck2001}, rather than direct observations.    

Let us assume that the measurement of the abundances of light elements
and the estimate of the anisotropy of cosmic rays at low energies may
be taken as realistic for the determination of the diffusion properties
of the ISM.

The anisotropy of cosmic rays at low energies is observed to be at the
level of $\delta \sim 10^{-4}$, and in the context of quasi-linear
theory (QLT) it is of order $v_D/c$, where $v_D$ is the drift velocity
of cosmic rays in the magnetic field of the Galaxy. 
The condition $v_D/c\sim 10^{-4}$ implies $v_D\sim 3\times 10^6$
cm/s. This is in good agreement with the theory again, because in 
QLT the streaming instability forces the streaming of cosmic rays to
occur at bulk velocities lower than the Alfven speed,
$v_A=B/\sqrt{4\pi\rho}\sim 2\times 10^{6}$ cm/s for $B=3\mu G$ and 
gas density $0.1\rm cm^{-3}$ (this should be considered as an average 
value over the magnetized halo of the Galaxy, say within 3 kpc from
the disk).  
In other words, the anisotropy is exactly what one would expect on 
the basis of bulk motion of cosmic rays at the Alfven speed
($v_D=v_A$). In QLT the pathlength for a particle to suffer a
change in direction by 90 degrees is 
\begin{equation}
\lambda = \frac{c}{\Omega \left(\frac{\delta B}{B}\right)^2},=
\frac{r_L(E)}{{\cal F}(k(p))}
\end{equation}
where $\Omega=c/r_L(p)$ is the gyration frequency of the particle
and $k=1/r_L(p)$. 

The pathlength $\lambda$ determines the confinement time in a region
of size $L$ as  
\begin{equation}
\tau = \frac{L^2}{c \lambda}.
\end{equation}
From observations of the abundance of light elements this time is
measured to be $\sim 3\times 10^6$ years, while from the abundance of
unstable radioactive isotopes one gets a
 larger number, $\sim
2\times 10^7$ years \cite{slibook}. These two numbers correspond respectively to
$\lambda=10$ pc and $\lambda=1.5$ pc. Here we assumed that the
magnetized region of the Galaxy in the direction perpendicular to the
disk has a typical size $L=3$ kpc. Note also that rigorously we may
use the parallel diffusion coefficient to estimate the escape from the
disk only if the magnetic field is oriented along $z$, which is at odds
with the conventional models of Galactic magnetic field. Therefore it
is worth keeping in mind that a more realistic estimate is likely to
differ from the one just illustrated and often used in the literature.   

From the equation for $\lambda$ one immediately obtains:
\begin{equation}
\epsilon_1=kP(k)=\left(\frac{\delta B}{B}\right)^2 = 3.5 \times 10^{-8}
\end{equation}
for $\lambda=10$ pc and 
\begin{equation}
\epsilon_2=kP(k)=\left(\frac{\delta B}{B}\right)^2 = 2.4 \times 10^{-7}
\end{equation}
for $\lambda=1.5$ pc. For the numerical evaluation we considered
cosmic rays with mean energy $1$ GeV. These values of $kP(k)$
correspond to $\delta B/B_0\sim 2\times 10^{-4}$ and $\sim 5\times
10^{-4}$ respectively on the relevant scales. On such scales it
appears that the assumptions of QLT are fulfilled. 

If the power spectrum is in the form of a power law, we can write
$P(k)=P_0 \left(\frac{k}{k_0}\right)^{-\alpha}$ and limit ourselves to
the two interesting cases $\alpha=5/3$ (Kolmogorov spectrum) and
$\alpha=3/2$ (Kraichnan spectrum). In both these cases most power is in
the form of modes with the largest spatial scale (namely at
$k_0$, assumed here to be $k_0\approx 1/100\pc$). The modes of
wavenumber $k_0$ resonate with particles with energy $E_0=2.8\times
10^{17}$ eV. The propagation of particles with energies larger than
$E_0$ is described in terms of a diffusion coefficient with a steeper
energy dependence than the one discussed here (Bohm diffusion) and
eventually straight line propagation. From the numerical values
obtained above, and assuming that $k_0\approx 1/100\pc$, one easily
infers that the power on a scale $k_0$ is  
\begin{equation}
P_0 k_0 \approx \epsilon_1 \left(\frac{k(1
  GeV)}{k_0}\right)^{\alpha-1} = 3.2\times 10^{-3} ~~~ (1.8\times 10^{-4})
\end{equation}
for $\lambda=10$ pc and $\alpha=5/3$ ($\alpha=3/2$), and 
\begin{equation}
P_0 k_0 \approx \epsilon_2 \left(\frac{k(1
  GeV)}{k_0}\right)^{\alpha-1} = 0.02  ~~~ (1.3\times 10^{-3})
\end{equation}
for $\lambda=1.5$ pc and $\alpha=5/3$ ($\alpha=3/2$).

These estimates show that the total power in the turbulent field may
be appreciably smaller than unity, which of course affects the
normalization of the diffusion coefficient, the confinement time and
the expected anisotropy at higher energies. The main problem with
these estimates is that they are based solely upon the parallel
diffusion coefficient, which, as discussed below may be incorrect. 
The issue of the strength of the turbulent field relative
to the strength of the regular field remains therefore open. 

It is worth stressing that
for $\alpha=5/3$ the diffusion approximation is broken at
$E_{th}\approx 8\times 10^{15}$ eV when $\lambda=10\pc$ and 
$E_{th}\approx 2\times 10^{18}$ eV when $\lambda=1.5\pc$. 
For $\alpha=3/2$ we have $E_{th}\approx 10^{14}$ eV 
for $\lambda=10$ pc and $E_{th}\approx 6\times 10^{15}$ eV for
$\lambda=1.5$ pc. This implies that at energy $E_{th}$ the anisotropy
is expected to become of order unity. Among all cases considered, the
only case that seems to be compatible with the fact that no large
anisotropy is observed up to the knee is the case $\alpha=5/3$ and
$\lambda=1.5\pc$. Note that this does not necessarily imply that a
large anisotropy should be observed at $E_{th}\approx 2\times 10^{18}$
eV, since at this energy the chemical composition in the Galaxy is
expected to be contaminated by heavy elements, which are as isotropic
as the particles with energy $E_{th}/Z$. Despite the interesting
conclusion, this has to be considered just as a hint, because of the
several assumptions that enter the previous estimate (for instance the
value of $L$ and $k_0$ and assumptions about geometry of the system).

The predicted escape time from the Galaxy as a function of energy is
more solidly predicted to be $\tau(E)\propto E^{-1/3}$ for Kolmogorov
spectrum and $\tau(E)\propto E^{-1/2}$ for Kraichnan spectrum. It is
worth stressing that this simple prediction, widely used in the
literature, completely neglects the possibility of perpendicular
diffusion or when it is not neglected, the assumption is adopted that
the scaling with energy of the perpendicular diffusion coefficient is
the same as for the parallel diffusion coefficient. Unfortunately, as
we show below, the role of perpendicular diffusion in the Galaxy is
likely to be crucial. 

\subsection{Drifts}
\label{sec:drifts}

Gradients in the modulus or orientation of the large scale field
$\boldsymbol B_0$
induce drift motions in the direction perpendicular to both 
the local field and its gradient. The drift velocity of the 
guiding center can be written as \cite{rossi70}:

\begin{eqnarray*}
  \boldsymbol{V_\perp}&=&
    \frac{c p}{Z e B_0}\Big\{
    \frac{1}{2} \sin^2\alpha
    \frac{\boldsymbol{B_0}\times\boldsymbol{\nabla}B_0}{B_0^2}+
    \cos^2\alpha\frac{\boldsymbol{B_0}\times
    [(\boldsymbol{B_0}\cdot\boldsymbol{\nabla})\boldsymbol{B_0}]}{B_0^3}
    \Big\}\\
    &=& c r_L \Big\{
    \frac{1}{2} \sin^2\alpha
    \frac{\boldsymbol{B_0}\times\boldsymbol{\nabla}B_0}{B_0^2}+
    \cos^2\alpha\Big[
    \frac{\boldsymbol{B_0}\times\boldsymbol{\nabla}B_0}{B_0^2}+
    \frac{\big(\boldsymbol{\nabla}\times\boldsymbol{B_0}\big)_\perp}{B_0}
    \Big]
    \Big\}\,,
\end{eqnarray*}
where $\alpha$ is the pitch angle of the particle.
The first term in this expression reflects the transverse gradient of
the field strength while the second term represents the effect of the
curvature of the field lines.

The above expression should be interpreted as the drift velocity averaged
over a gyration period of the particle. As an estimate of the order of
magnitude of the time scale for escape from the region of size $L$ due
to drift motion, we can write $\tau_D(E)\sim \frac{L\lambda_{grad}}{c
r_L(E)}$, where $\lambda_{grad}$ is the spatial scale on which the
gradient in the magnetic field appears. This expression clearly shows
that if the drifts are relevant at all this may happen only at very high
energies.  

Three toy models are particularly interesting as far as drifts are
concerned and will be discussed in detail in \S \ref{sec:toy2}, \S
\ref{sec:toy3} and \S \ref{sec:toy4}. Here we limit our discussion to
the expected effects of drift motions. The first model (Toy model II
in \S \ref{sec:toy3}) has only spatially constant (in modulus)
azimuthal magnetic field. In this case the field lines are simply
concentric circles in $z=\tu{const.}$ planes. The only gradient is due to
the curvature of the magnetic field lines and the drift velocity is
given by 
\begin{equation}
  \boldsymbol{v_D} =  E_{18}\,c\,\cos^2\alpha\,\frac{\boldsymbol{\hat{z}}} {\rho},
\label{eq:drift1}
\end{equation} 
where $\hat z$ is the unit vector in the $z$ direction, $\rho$ is the
distance (in kpc) from the center in the plane $z=0$ and $E_{18}$ is the
particle energy in units of $10^{18}$ eV. Clearly this expression and
the ones we will list below are valid as long as the spatial scale of
the gradient is much larger than the Larmor radius of the particles.
This condition also assures that the drift velocity is always smaller
than the speed of light. From Eq. \ref{eq:drift1} one can see that the
drift pushes the particles perpendicular to the plane.

The second toy model that we will consider is similar to the previous
one but with the strength of the magnetic field having a gradient
along the $\rho$ direction (see Eq. \ref{eq:Brad}). 
It is easy to predict that also in this
case the drift velocity is oriented along the $\hat z$ direction. The
drift velocity in this case is 
\begin{equation}
  \boldsymbol{v_D} =
  E_{18}\,c\,\frac{1}{17}(1+\cos^2\alpha)\,{\boldsymbol{\hat{z}}}~~~~~ \rm for~ \rho>4 kpc.
\label{eq:drift2}
\end{equation}
Finally, in the third toy model we assume that the magnetic field is
still azimuthal but is constant in the $z=0$ plane and has a gradient
in the $\hat z$ direction (see Eq. \ref{eq:Bzeta}). In this case
the drift velocity is 
\begin{equation}
  \boldsymbol{v_D} =   E_{18}\,c\,\Big[\cos^2\alpha\,\frac{1}{\rho}
  {\boldsymbol{\hat{z}}}+
  \sin^2\alpha\,\frac{1}{2 z_c} {\boldsymbol{\hat{\rho}}}\Big]\exp(z/z_c). 
\label{eq:drift3}
\end{equation}
Clearly in this third case the direction of the drift is no longer
along $\hat z$ and depends on 
$\rho$. 

\section{Determination of the parallel and perpendicular diffusion
  coefficients} \label{sec:diffu}

To calculate the diffusion coefficients we inject a few thousand
particles of a given energy isotropically in a magnetic field composed
of a constant regular component along $\hat{z}$ plus a uniform turbulent
component. We record the particle trajectories and we then calculate the
instantaneous parallel and perpendicular diffusion coefficients as:
\begin{equation}\label{eq:de}
  D_\parallel(\tau)=\frac{\langle \Delta z^2\rangle}{2 \tau}
  \qquad\tu{and}\qquad
  D_\perp(\tau)=\frac{\langle \Delta x^2\rangle}{2 \tau}=\frac{\langle \Delta
  y^2\rangle}{2 \tau}\,.
\end{equation}

\begin{figure}
  \centering
  \includegraphics[width=0.45\textwidth]{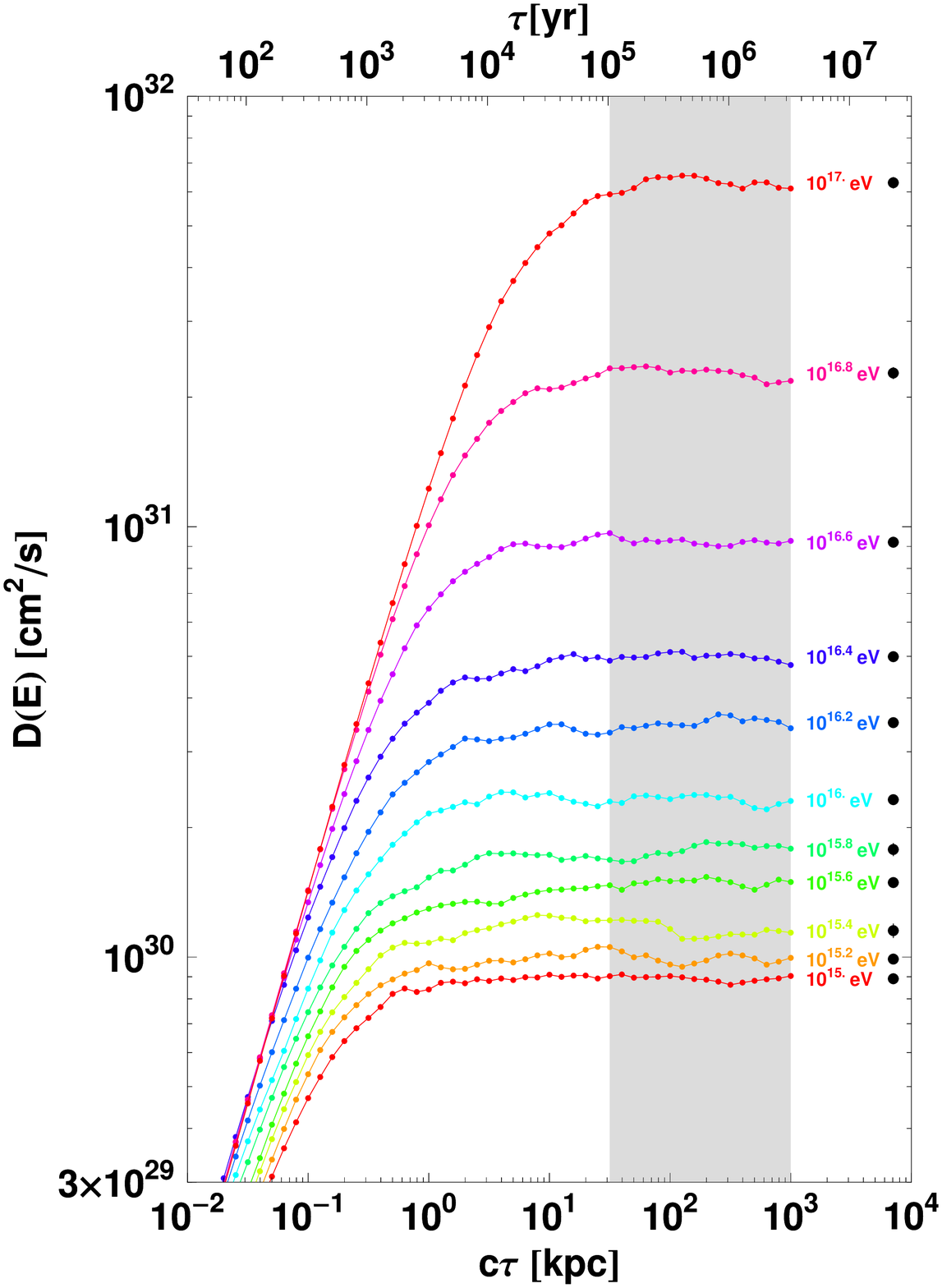}\quad
  \includegraphics[width=0.45\textwidth]{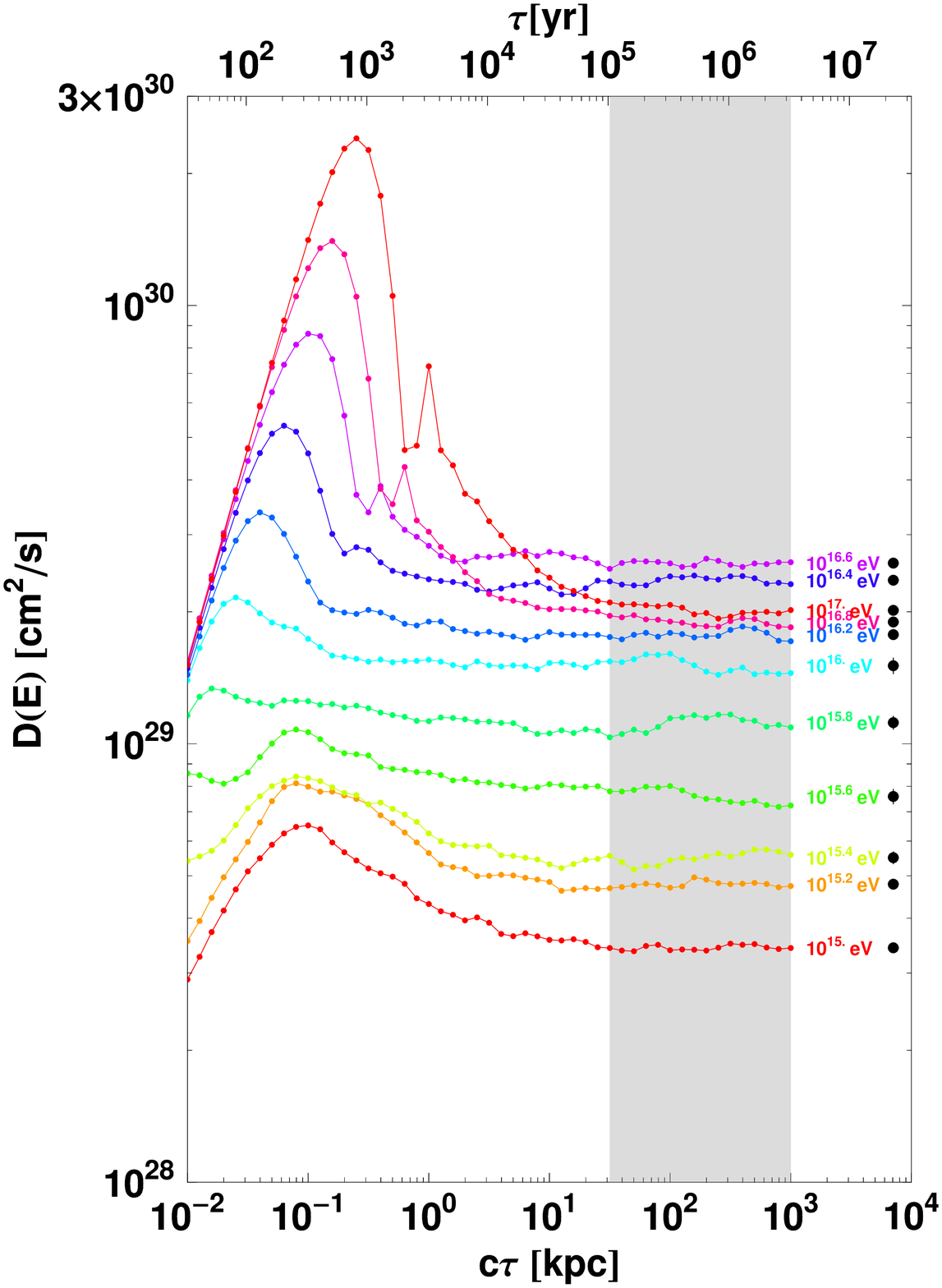}
  \caption{Instantaneous diffusion coefficients as a function of
  propagation time. Left: parallel, right: perpendicular. Each
  line/color corresponds to a different particle energy as indicated in
  the plot. The black points on the far right of each line indicate the
  average of the corresponding points in the gray region and they
  represent the estimate of the diffusion coefficient at the
  corresponding energy. Note the different scales on $y$-axes.
  } \label{fig:idc}
\end{figure}
We plot the instantaneous diffusion coefficients as a function of
propagation time in Fig.~\ref{fig:idc} for the case $\delta B /B_0 =1$.
The left panel is the parallel diffusion coefficient while the right one
is the perpendicular diffusion coefficient. 
The parallel instantaneous diffusion
coefficient increases linearly in the beginning when the particles are
still only feeling the regular field and at some point flattens when the
full diffusive regime is reached, typically within a few scattering
lengths. In the perpendicular case the instantaneous diffusion
coefficient increases for a time $\tau_\tu{L}/2\simeq\pi\times
r_\tu{L}/c$, corresponding to half a gyration around the regular field.
At this point continuing the gyration the particle is going back to its
starting position and the diffusion coefficient is decreasing, having a
minimum at $\tau_\tu{L}$. After some time the diffusion regime is
reached and the curve shows a plateau. This plateau identifies the
diffusion coefficient and we use the average of the last 15 points (the
gray region in the plots) to estimate it.

In the following few paragraphs we present our results for the diffusion
coefficients as a function of energy for some interesting
configurations.

\subsection{The case of vanishing regular field}\label{sec:de3d}
\begin{figure}
  \centering
  \includegraphics[width=0.9\textwidth]{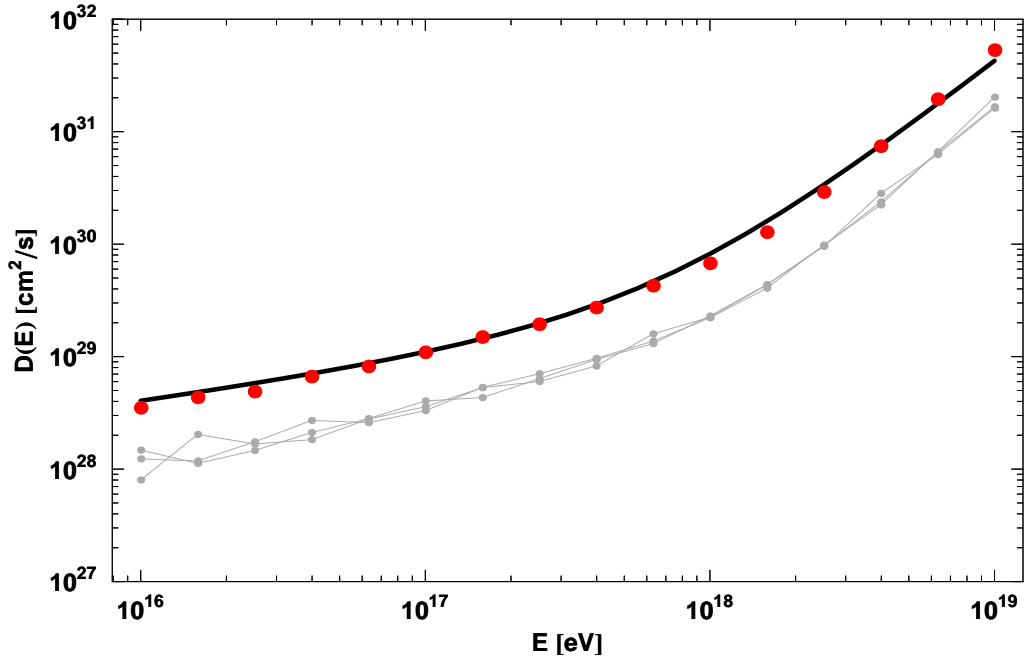}
  \caption{Diffusion coefficient for a configuration with vanishing
  regular field. The gray points and lines are the diffusion
coefficients along the three axes, the red points are the total diffusion
coefficient and the black line is the parametrization of the diffusion
coefficient given in Ref.~\cite{parizotdiffu}.}\label{fig:3d}
\end{figure}

Without a background field the only type of turbulence that can be
considered is isotropic turbulence. In Fig.~\ref{fig:3d} we plot the
diffusion coefficient as a function of energy for 3D turbulence in a
configuration with no regular field, but only turbulent field with
$L_\tu{max}=100\pc$ and $\delta B=100\muG$\footnote{Please note that
here and in the following when denoting $\delta B=100\muG$ we actually
mean: $\sqrt{\langle\delta B^2\rangle}=100\muG$.}.

In this case, in order to compare our results with the ones of
Ref.~\cite{parizotdiffu}, we calculated the diffusion coefficients using
$6$ in the denominator of Eq. (\ref{eq:de}) instead of $2$. In
Fig.~\ref{fig:3d} the gray points and lines are the diffusion
coefficients along the three axes, the red points are the total diffusion
coefficient and the black line is the parametrization of the diffusion
coefficient given in Ref.~\cite{parizotdiffu} that was obtained from
simulations using the plane wave approach. In this case we used the FFT
approach and the agreement is very good.

\subsection{Combination of regular and turbulent fields}
\begin{figure}
  \centering
  \includegraphics[width=0.9\textwidth]{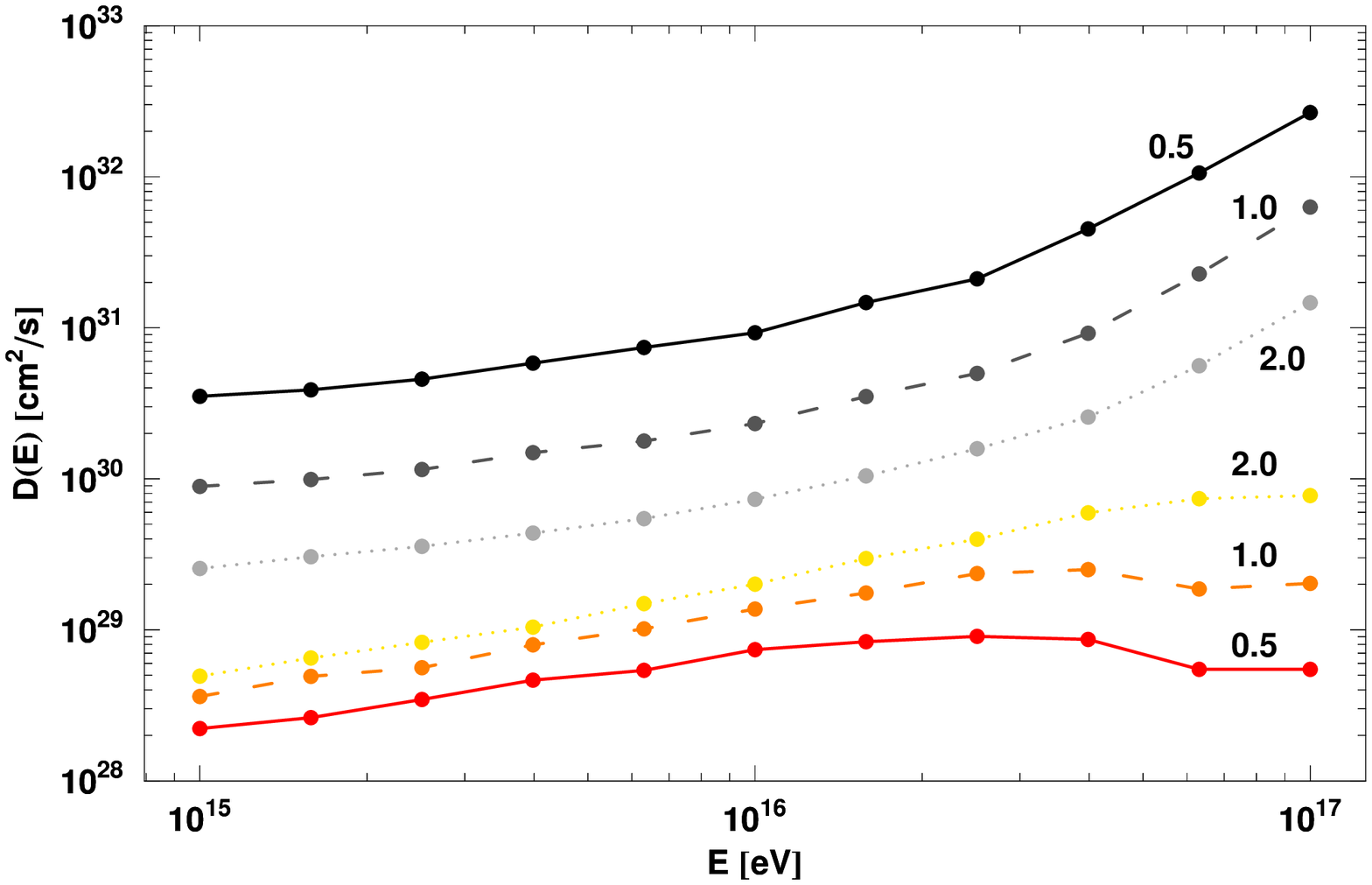}
  \caption{Parallel and perpendicular diffusion coefficients as a
  function of energy for three levels of turbulence. The upper three
  lines are the parallel diffusion coefficients, while the bottom three
  represent the perpendicular one. The level of turbulence, $\delta
  B/B_0$ is given by the numbers attached to the lines.
  }\label{fig:d_e}
\end{figure}

In this case we use a superposition of a constant background field and a
turbulent field with three levels of isotropic turbulence: $\delta
B/B_0=0.5,1,2$. The maximum scale of the turbulence is set to
$L_\tu{max}=0.1\kpc$, $B_0=1\muG$ and we use the FFT approach to
generate the turbulence. We plot the parallel and perpendicular
diffusion coefficients in Fig.~\ref{fig:d_e}. The top three lines
represent the parallel diffusion coefficients, while the bottom three
the perpendicular ones. The turbulence level is given by the numbers
attached to the curves. It is interesting to note that while the low
energy ($10^{15}$- $10^{16}$ eV) slope of the parallel diffusion
coefficient is $1/3$ as one would expect, the slope of the
perpendicular one is steeper, being about $0.5-0.6$. 

\begin{figure}
  \centering
  \includegraphics[width=0.9\textwidth]{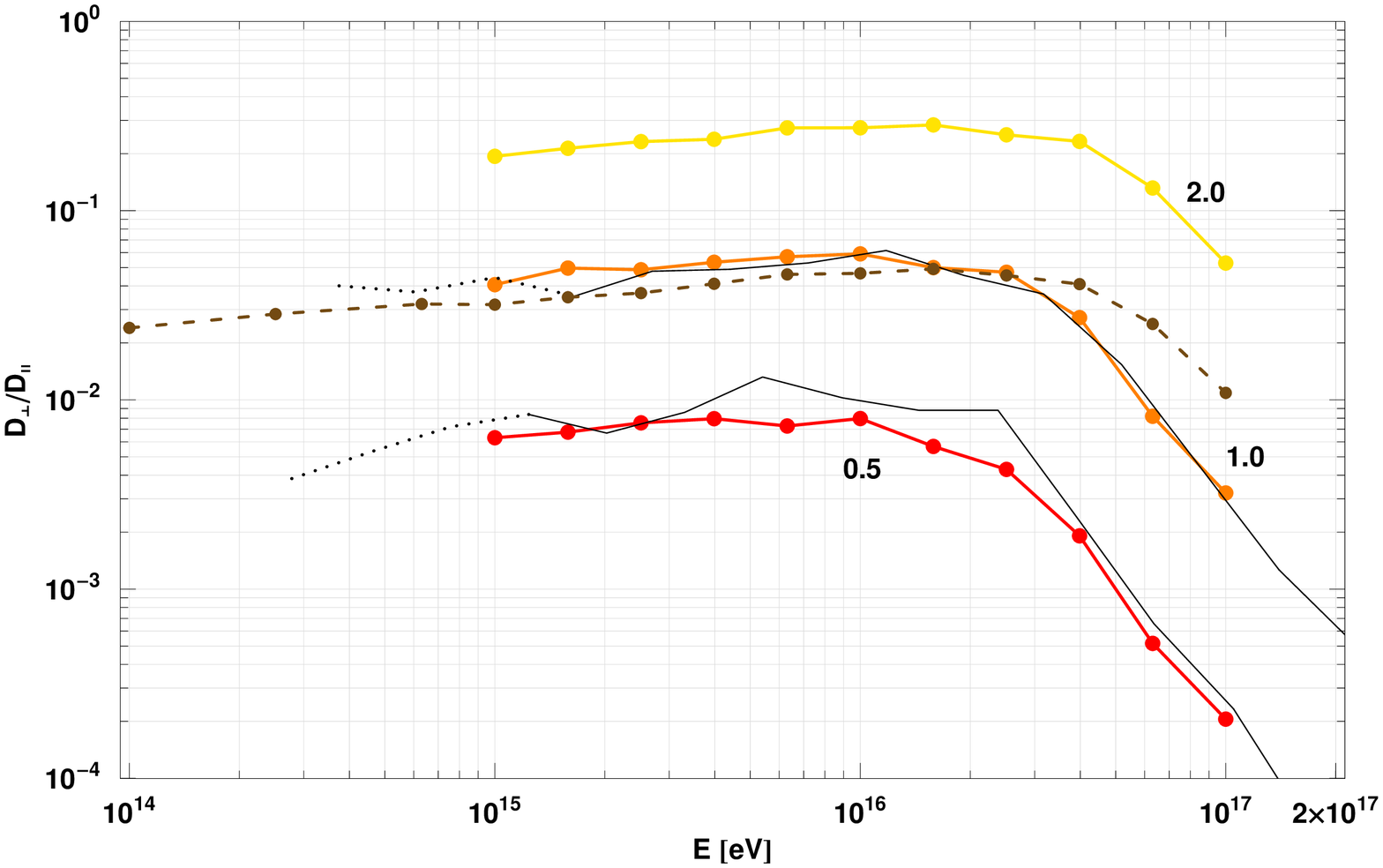}
  \caption{Ratio of the perpendicular to parallel diffusion
  coefficients, $D_\perp/D_\parallel$, as a function of the energy. The
  three set of points connected by solid lines are the results
  for the three levels of turbulence indicated. The points connected by
  the dashed line are the result of a simulation with a set of
  parameters similar to the one used for the orange one, but in this
  case using the plane waves approach instead of the FFT one.
  The two black thin lines are the results of Ref.~\cite{clp} for
  $\delta B/B_0=0.92$ (upper one) and $0.52$ (lower one).
  }\label{fig:dpdp_e} 
\end{figure}

In Fig.~\ref{fig:dpdp_e} we plot the ratio of the perpendicular to
parallel diffusion coefficients, $D_\perp/D_\parallel$, as a function of
energy. The three sets of points connected by solid lines are the
results of the three simulations shown in the previous plot. We compared
our results with the ones obtained in Ref.~\cite{clp}. The thin black
lines are the results from their Fig.~6 for the cases $\eta=0.46$ and
$0.21$ that correspond to $\delta B/B_0=0.92$ and $0.52$ respectively.
The agreement between the two sets of results is pretty good, especially
for the case $\delta B/B_0 \simeq 1$.
 The ratio of the diffusion coefficients is almost constant with
 a slow $E^{(0.1-0.2)}$ energy dependence. 

The tiny difference in slope between $D_\perp$ and $D_\parallel$ at low
energy is more apparent in this plot. It is interesting to note that
this difference seems to be present also in the results of Ref.~\cite{clp},
at least for the case $\delta B/B_0 \simeq 1$. In the case $\delta B/B_0
\simeq 0.5$ the scattering of their points is too big to allow for
inferring any conclusions in this respect.
In order to confirm that this slope was not a systematic
effect due to the method used to generate the turbulence, we performed a
simulation using the plane wave approach to generate the field. These
results are represented by the brown dashed line. The simulation
parameters and the shape of the turbulence spectrum in this case are a
bit different from the others, and the resulting curve does not coincide
with the one obtained from the FFT approach, but also in this case the
ratio is not constant at low energy and presents a small positive slope.

The results of Ref.~\cite{GJ} show no dependence of the ratio
$D_\perp/D_\parallel$ on energy and for $\delta B/B_0 = 1$ their
result is smaller than ours by about a factor 2.

\section{Toy models of the Galactic magnetic field}
\label{sec:toy}

The large scale structure of the Galactic magnetic field is likely to
be complex, as made of spiral arms and various types of gradients
along the radial direction in the disk and along the $\hat z$ axis,
perpendicular to the disk. The same presence of the spiral arms
induces gradients on different spatial scales. On top of this large
scale structure a turbulent component is present which turns out to be
responsible for the diffusive motion of cosmic rays. In all cases
presented below, the values of the quantity $\delta B/B_0$ is assumed to
be spatially constant (in other words the turbulent field is a
constant fraction of the large scale field). It appears rather
unrealistic that the naive expectations based on quasi-linear theory
may find an easy confirmation with this complex structure of the
magnetic field and indeed we confirm that this is in general not the
case. In order to understand the various reasons why the expectations
of QLT may be not fulfilled, in the following we discuss in detail
four toy models of the magnetic field of the Galaxy in both its
regular (large scale) and turbulent components. The first model is
that of a magnetized homogeneous sphere with only turbulent field. 
In this case QLT cannot even be applied because of the absence of a
regular field which does not allow to develop a perturbative
approach to particle propagation. In this case however the confinement
time that is obtained from simulations is close to the naive
extrapolation of QLT to a regime in which it should not be applied. 

The second toy model consists of a purely azimuthal, spatially
constant magnetic field. The particles are injected at the position of
the Earth and collected on the surface of a cylinder of radius 10 kpc
and height 0.5 kpc. 

The third and fourth toy models are variations of the previous one
with the addition of gradients along the radial direction and along
the $z$ direction. 

\subsection{Toy model I: a magnetized homogeneous sphere}
\label{sec:toy1}
\begin{figure}
  \centering
  \includegraphics[width=0.9\textwidth]{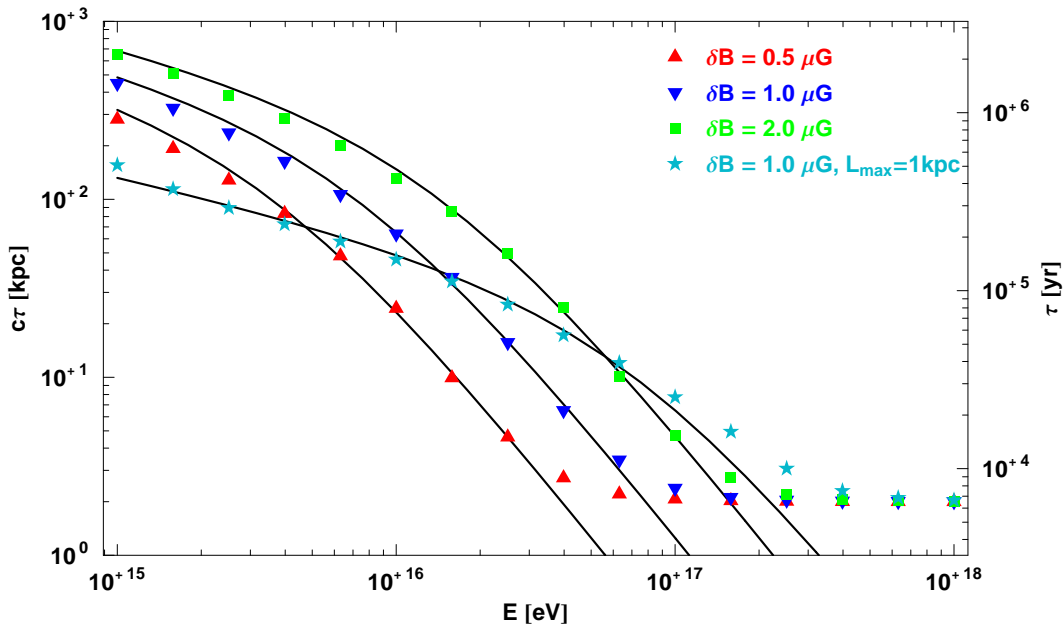}
  \caption{Times of escape from a sphere filled with uniform turbulent
  field for protons injected in the center. The levels of turbulence are
  indicated in the plot. The first three cases are for
  $L_\tu{max}=0.1\kpc$, while the last one for $L_\tu{max}=1\kpc$. The
  black lines are the expected results obtained using the diffusion
  coefficient given in Ref.~\cite{parizotdiffu}.
  }\label{fig:sphere}
\end{figure}

We consider a sphere filled uniformly with isotropic turbulent field
with $L_\tu{max}=0.1\kpc$ and $\delta B = 0.5, 1, 2\muG$. We inject
protons in the center of the sphere and we collect them when they reach
a distance of $2\kpc$ from the center. The times of escape from the
sphere are plotted as triangles and boxes in Fig.~\ref{fig:sphere}. We
also plotted, with stars, the results obtained using $L_\tu{max}=1\kpc$
instead of $L_\tu{max}=0.1\kpc$ for the case $\delta B = 1\muG$. The
black lines are the expected propagation times obtained using the
parametrization of the diffusion coefficient given in
\cite{parizotdiffu} and already used in \S \ref{sec:de3d} for
comparison:
\begin{equation}
  \tau(E) = \frac{R^2}{6 D(E)}\,.
\end{equation}
The agreement is very good both in the low energy and in the transition
region. Going to very high energies, the transition to straight line
propagation becomes visible.

\subsection{Toy model II: large scale azimuthal field with no spatial
  gradients} 
\label{sec:toy2}

\begin{figure}
   \centering
   \includegraphics[width=0.7\textwidth]{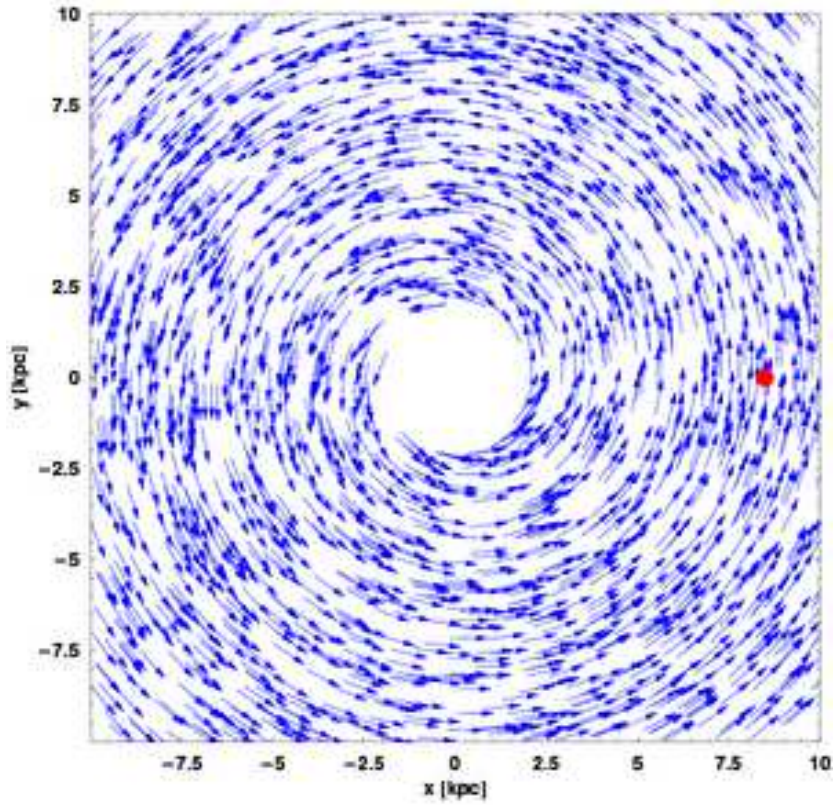}
   \caption{Azimuthal magnetic field in Toy model II.}\label{fig:Baz} 
\end{figure}

\begin{figure}
   \centering
   \includegraphics[width=0.9\textwidth]{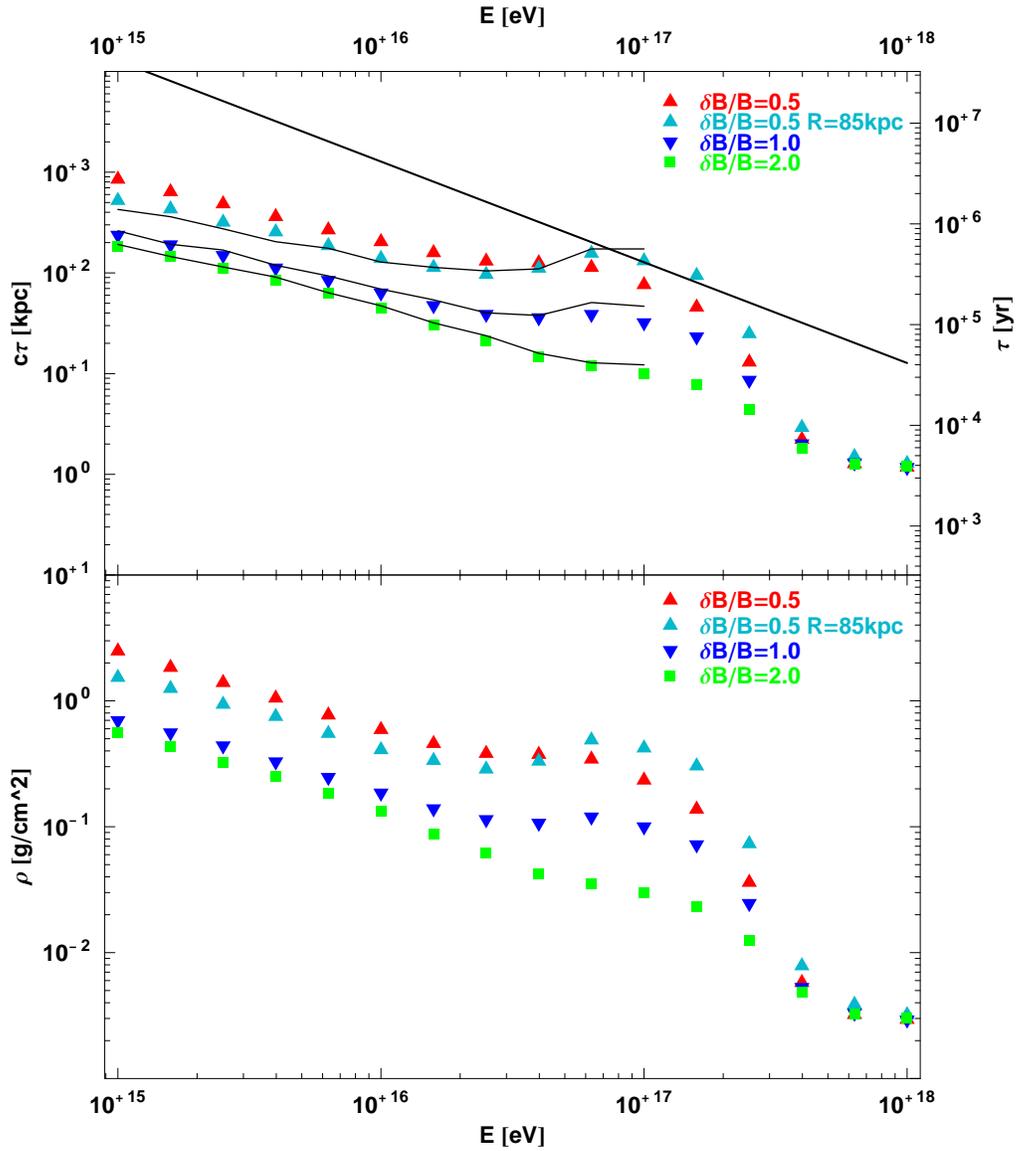}
   \caption{Particle escape times for Toy Model II. The upper panel
   shows the times required for the particles to escape from a cylinder
   with half-height of $0.5\kpc$. The lower panel shows the grammage of
   gas crossed. The boxes and triangles are the values for different
   levels of turbulence as indicated in the plot. The injection is set
   at $8.5\kpc$ for all cases except for the light blue upward triangles
   for which it is set at $85\kpc$. The thick black line is the drift
   timescale while the thin black curves represent the diffusion
   timescales.}\label{fig:Baz_times}
\end{figure}

The magnetic field as seen from above the disk is as shown in
Fig. \ref{fig:Baz}. 
This field structure is assumed to resemble at 
least qualitatively the spiral structure of the Galactic field. In
passing we notice that this purely azimuthal field has also recently
been adopted by \cite{hora}. The turbulent field is assumed to have a
Kolmogorov spectrum with a largest scale $L_{max}=0.1$ kpc and total
power $\delta B/B_0=0.5,1$ and $2$. Particles are injected at the
Earth, located at $R_\odot=8.5$ kpc from the center and propagated backwards 
in time until they escape the cylinder of radius $R=10$ kpc and height
above and below the disk of $0.5$ kpc. A crucial point to realize here
is that the magnetic field lines are closed loops: the magnetic field
strength is spatially constant but the orientation of the field
 changes as illustrated in Fig. \ref{fig:Baz}. The fact that the field
lines are closed implies a straightforward but important conclusion:
the particles cannot escape the cylinder by diffusing parallel to the
magnetic field lines. The only way particles can escape is by
diffusing and drifting perpendicular to the field lines, which is
clearly made more difficult by the smallness of the perpendicular
diffusion coefficient (see \S \ref{sec:diffu}) as compared with the
parallel  diffusion coefficient. The escape times of cosmic rays 
as functions of energy for the various cases that have been
calculated are illustrated in Fig. \ref{fig:Baz_times} (top panel).
The lower panel of the figure illustrates the column densities
experienced by cosmic rays with given energy. The gas density has been
assumed to be constant and equal to $1\cm^{-3}$ inside the disk
($|z|<200$ pc) and $0.01\cm^{-3}$ outside the disk. The different
symbols refer to the values of $\delta B/B_0$ as indicated in the figure.
The straight line represents the drift time calculated from Eq.
(\ref{eq:drift1}) using the average drift velocity. It is worth noticing
that the actual drift times have a very extended tail towards 
large times, due to the dependence of this quantity on the angle of
injection of the particles with respect to the large scale local field.
The black lines and dots are the diffusion time scales, $\propto
D_\perp(E)^{-1}$, where the perpendicular diffusion coefficient is taken
from the simulations described in \S \ref{sec:diffu}.

A general comment about the relative role of parallel and perpendicular
diffusion is in order: parallel diffusion is more effective than
perpendicular even in the case of strong turbulence considered here,
but it only leads to motion of the particles along the closed magnetic
field lines. Perpendicular diffusion, though much slower, is
responsible for particle escape in the direction perpendicular to the
disk (there is also some escape from the sides of the cylinder but
this process is less efficient because the sides are $\sim 1.5$
kpc away from the location of the Earth, while the halo has been
assumed to be only $0.5$ kpc thick). The parallel with the Galaxy is very
instructive in this instance: particles diffuse effectively along the
spiral arms, whose length is roughly $R_{arm} \sim \pi R_\odot \sim
30$ kpc long (at the distance of the Sun from the galactic center). 
The diffusion time parallel to the arms is therefore
$\tau_\parallel\sim R_{arm}^2/D_\parallel$. At the same time, cosmic
rays diffuse in the direction perpendicular to the disk in a time 
$\tau_\perp \sim R_H^2/D_\perp$. The ratio of the two time scales is
$\tau_\parallel/\tau_\perp \sim 10^3 D_\perp/D_\parallel$, where we
assumed $R_H\sim 1$ kpc. For $\delta B/B_0\sim 1$ the perpendicular
diffusion coefficient is not much smaller than $D_\parallel$, so that
it is easy to understand that perpendicular diffusion may become the
dominant channel of cosmic ray escape from the Galaxy rather than
parallel diffusion. In our toy model this situation is extreme in that
the magnetic field lines are closed and no escape at all is possible
along the field. As a consequence, the energy dependence which is
illustrated in Fig. \ref{fig:Baz_times} reflects the energy dependence
of the perpendicular diffusion coefficient, which in the relevant
energy range can be approximated as $D_\perp \propto E^{0.5-0.6}$. It is
instructive that such a slope, usually associated (at low energies) to
a Kraichnan spectrum of turbulence (parallel diffusion) can in fact be
achieved with a Kolmogorov perpendicular diffusion (at least in the
high energy range we are able to treat here).

The important role of perpendicular diffusion in determining the
escape time is also shown by the absolute normalization of the curves
in Fig. \ref{fig:Baz_times}. For parallel diffusion, at least in the
quasi-linear regime, one expects the diffusion coefficient to
decrease while increasing $\delta B/B_0$, so that the escape times
increase. In our toy model the perpendicular diffusion coefficient in
fact increases with increasing $\delta B/B_0$.

Aside from diffusion, the escape times are also affected by drift
motions. In particular, drifts become important where the drift time
(the straight line in Fig. \ref{fig:Baz_times}) becomes of the same
order of magnitude of the diffusion times (black lines). For $\delta
B/B=0.5$ this happens at energies around $10^{17}$ eV, while drift seems
irrelevant for stronger levels of turbulence. Besides this effect, which
is rather clear from Fig. \ref{fig:Baz_times}, there is a more subtle
effect induced by drifts, which is evident in the low energy part of the
curve for $\delta B/B_0=0.5$. One can notice that the black line
illustrating the effect of diffusion (for $\delta B/B_0=0.5$) lies below
the upward red triangles obtained in the simulation. In order to
understand the reason for this apparent problem, we calculated the
escape times in the case in which the Earth is located at 85 kpc from
the center instead of 8.5 kpc. In this case the cylinder is larger but
it has the same height. But more important the curvature of the magnetic
field lines is reduced appreciably so that the drift velocity drops
correspondingly. One can see that the low energy behaviour in this
case (upward light blue triangles) agrees well with the black curve,
therefore confirming that the reason for the slim disagreement has
something to do with the curvature of the field lines.

To achieve a better understanding of the modifications the drifts
produce to the diffusion process we calculated the diffusion
coefficients in this geometry and we found that the drifts are modifying
the two perpendicular diffusion coefficients reducing the one along $z$
and increasing the one along $\rho$. In fact one should keep in mind
that the concepts of parallel and perpendicular diffusion were
introduced here with reference to the specific case of a large scale
coherent background field, with no intrinsic curvature of the field
lines. When the field lines are curved, then the definition itself of
parallel and perpendicular diffusion changes, as discussed in detail
in \ref{appendix}.

At energy $\sim 10^{17}$ eV the Larmor radius of the particles equals
the maximum wavelength in the power spectrum of the
turbulent field and the diffusion regime changes, gradually shifting
toward the straight line propagation, which in Fig. \ref{fig:Baz_times}
corresponds to the extreme right, flat part of the curves for the escape
time. It is worth reminding the reader that all these considerations
remain valid for heavier nuclei once the energy is substituted by
rigidity. 

We conclude this discussion of the second toy model with a comment on
the absolute magnitude of the escape time. Though keeping in mind that
this is a toy model of the magnetic field of the Galaxy, we believe
that some qualitative conclusions can be drawn. At energy $10^{15}$ eV
the escape time for the cases considered here is $\tau_{15}\approx
0.5-5$ million years (the halo height here is only $0.5$ kpc). These
numbers are of the same order of magnitude of the confinement times
estimated from the abundance of light element in the $GeV$ region,
which means that in order to fit these observations one should postulate
that the escape time below $10^{15}$ eV should be practically energy
independent. We could not envision any realistic mechanism able to
justify such an expectation. It follows that within the limitations of
the toy model 2 it is very hard to obtain a realistic, even
qualitative, description of what is observed in the Galaxy at much
lower energies. This conclusion is confirmed also by the curves on the
grammage: at $10^{15}$ eV cosmic rays traverse a column
density of $1-2~\rm g~cm^{-2}$. As we discuss below, this conclusion
is the same for the other toy models considered here. 

\subsection{Toy model III: large scale azimuthal field with spatial
  gradient along $\hat \rho$} 
\label{sec:toy3}

\begin{figure}
  \centering
  \includegraphics[width=0.9\textwidth]{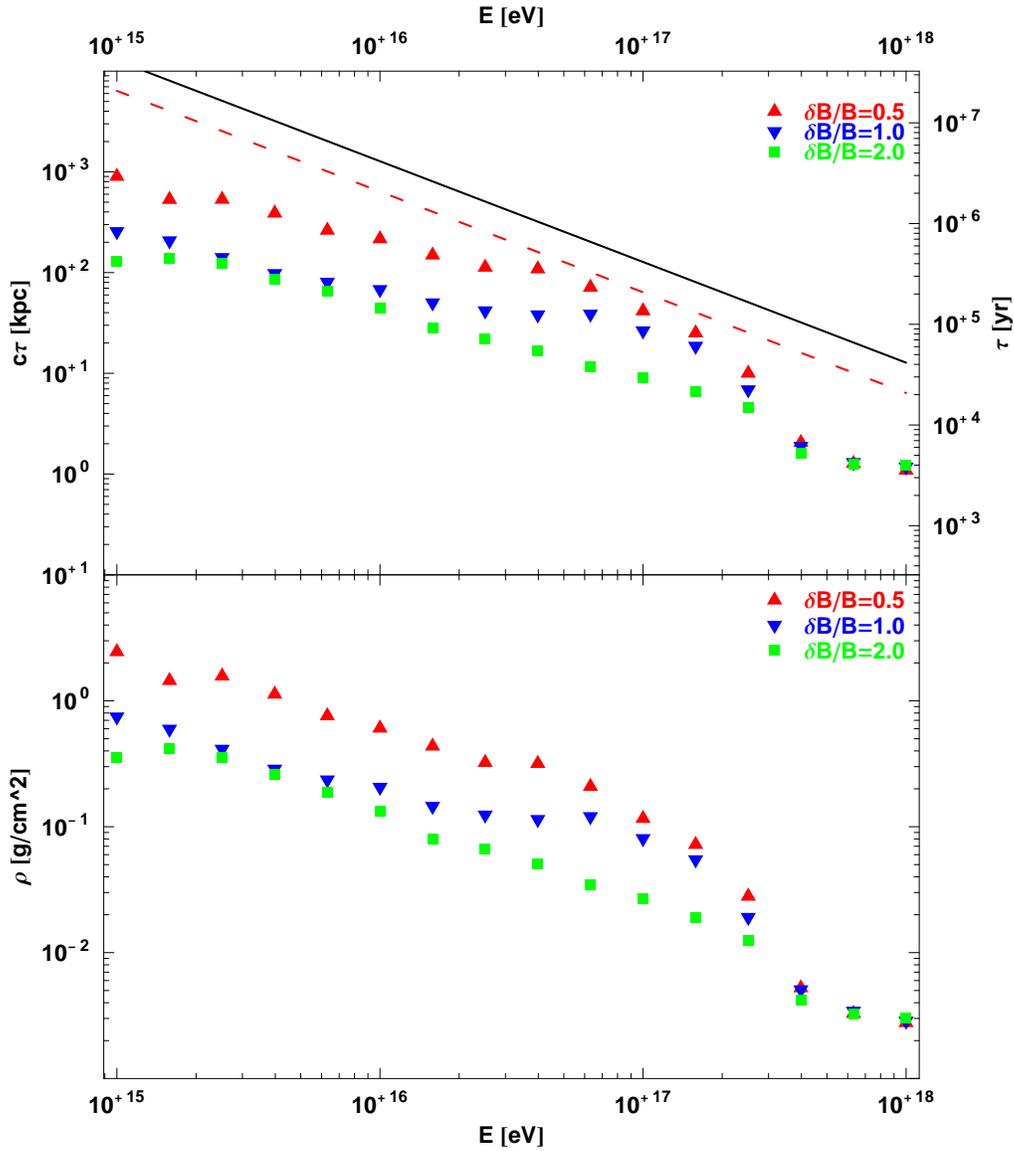}
  \caption{Particle escape times for Toy Model III. The black line is
  the drift timescale for the field of Toy Model II, while the red
  dashed line is the drift timescale for the present
  configuration.}\label{fig:Baz_rho_times} 
\end{figure}

The global structure of the large scale azimuthal field is not changed
with respect to Toy Model II, but we introduce here a gradient of the
modulus of the large scale field with the radial coordinate $\rho$
measured in the $x-y$ plane. The radial dependence of the field is
assumed to be in the form:
\begin{displaymath}
  B(\rho)=\left\{
\begin{array}{l l}
2.125 \mu G & \rho<4 \\
8.5\mu G ~ \rho^{-1} & \rho>4
\end{array}
\right.\,,
\label{eq:Brad}
\end{displaymath}
where $\rho$ is the radius in cylindrical coordinates in units of
kpc. As discussed in \S \ref{sec:drifts}, in this case the drift
velocity is still oriented in the $\hat z$ direction, therefore the
drift due to a gradient of the strength of the field behaves
qualitatively as the gradient due to the curvature in the field lines,
discussed in the section above. The escape time and the grammage for
this case are illustrated in Fig.~\ref{fig:Baz_rho_times}, where the
red dashed line indicates the drift timescale, again calculated using
the average drift velocity. At the distance of the Earth the gradient
due to the radial dependence reduces the drift time by roughly a factor
2, thereby making the line for the drift time scale almost touch the red
triangles ($\delta B/B_0=0.5$). For stronger levels of turbulence the
drifts become basically irrelevant, even at the highest energies. The
absolute normalizations of the time scales are affected very little by
the radial gradient of $\boldsymbol{B}$, therefore most comments made
for Toy Model 2 are valid here too. 

\subsection{Toy model IV: large scale azimuthal field with spatial
gradient along $\hat z$} 
\label{sec:toy4}

We conclude this section by investigating the case of an azimuthal field
with a gradient along $\boldsymbol{\hat z}$, as described by the
following expression:
\begin{displaymath}
  B(z)=\exp(-z/z_c) \mu G,
\label{eq:Bzeta}
\end{displaymath}
with $z_c=0.25$ kpc or $0.1$ kpc. In this case the drifts due to the
$z$-dependence are in the radial direction and, at the Earth position,
are bigger than the drifts due to the curvature of the field
lines. The sum of the two drifts tends to push the particles toward
the center of the Galaxy, where the drifts due to curvature dominate. 
The exit points of the particles in this case are shifted in the
direction of the galactic center, while in the previous two Toy Models
most particles escaped from a ring with $\rho\sim 8.5$ kpc. 

\begin{figure}
  \centering
  \includegraphics[width=0.9\textwidth]{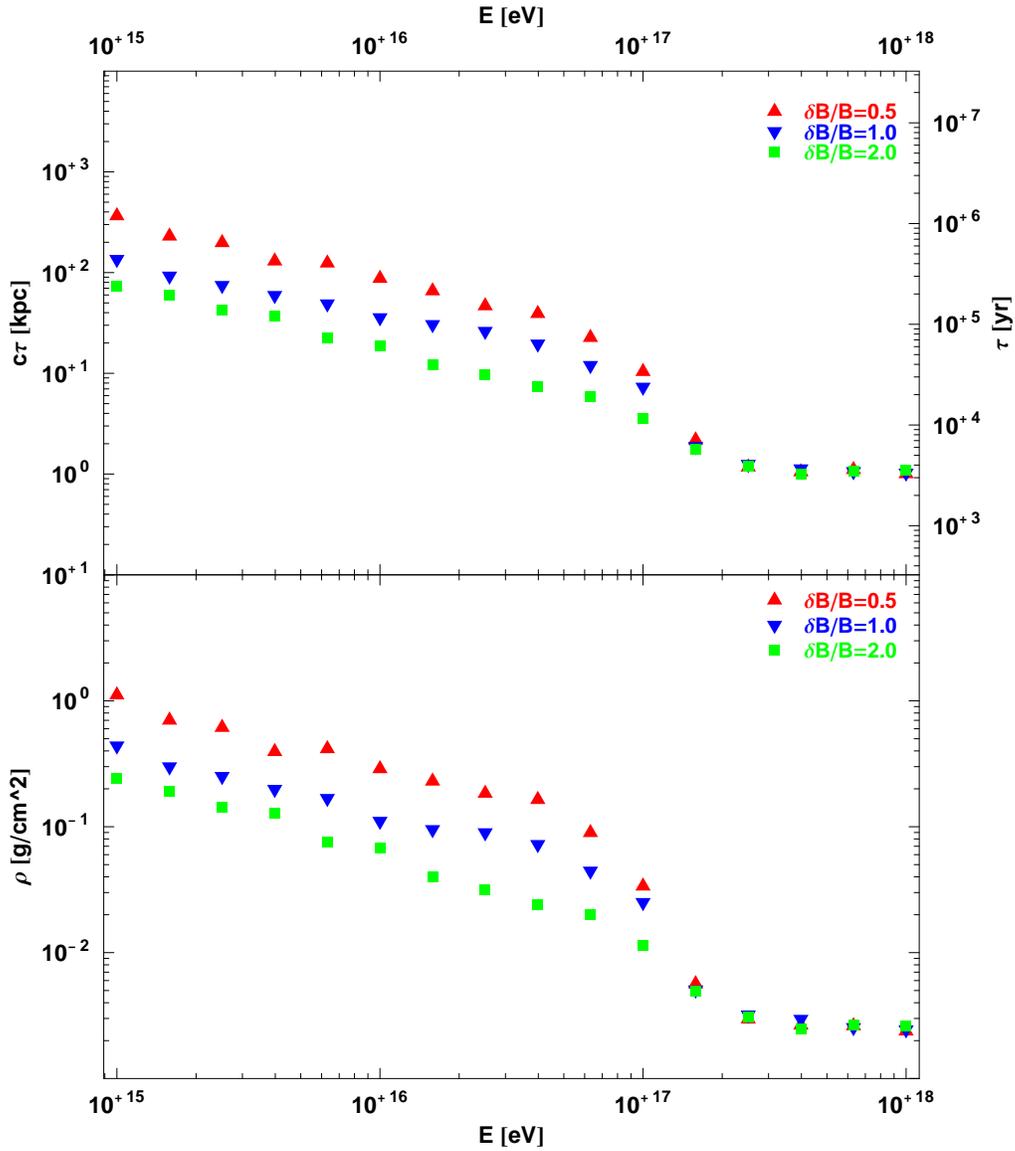}
  \caption{Particle escape times for Toy Model IV and $z_c=0.25\kpc$.}
  \label{fig:Baz_z025_times} 
\end{figure}

The escape times and grammage for $z_c=0.25$ kpc are illustrated in 
Fig.~\ref{fig:Baz_z025_times} with the usual meaning of the symbols. 

\begin{figure}
  \centering
  \includegraphics[width=0.9\textwidth]{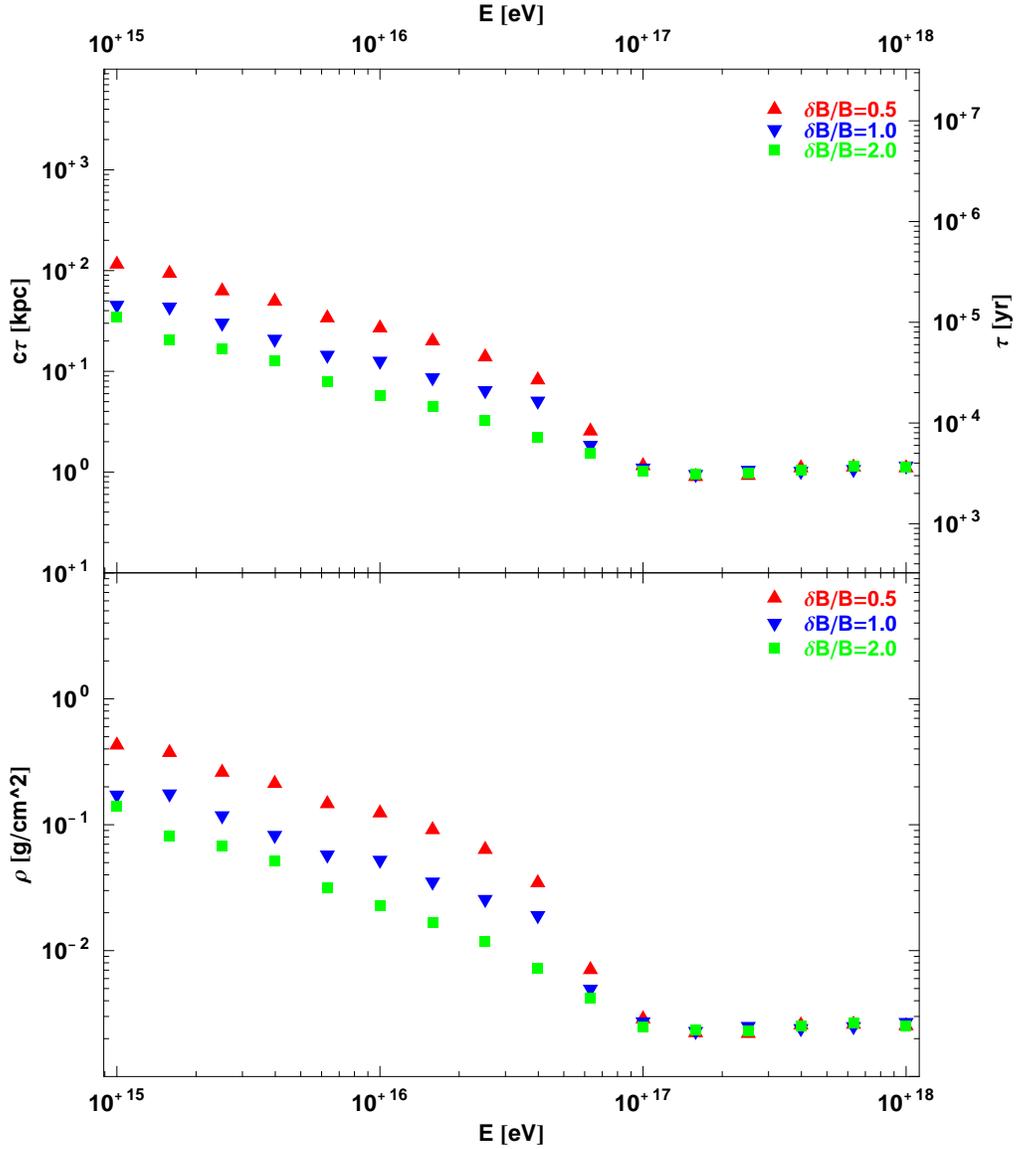}
  \caption{Particle escape times for Toy Model IV and $z_c=0.1\kpc$.}
  \label{fig:Baz_z01_times} 
\end{figure}

The effect of drifts, combined with the smaller effective size of the
magnetized halo along $\boldsymbol{\hat z}$, contribute to reduce the
escape times. At $10^{15}$ eV the escape time is always shorter than 1
million year. However the slopes of the curves, although rather
uncertain, do not seem to point toward any flattening that may help
reconcile the grammage at $10^{15}$ eV ($0.2-1~\rm g~cm^{-2}$) with that
observed in the GeV region. A further reduction of the escape times is
achieved by reducing the scale $z_c$. For instance the time scales and
grammage for $z_c=0.1$ kpc are illustrated in Fig.~\ref{fig:Baz_z01_times}. 

\section{Discussion and Conclusions}\label{sec:discussion}

The propagation of cosmic rays in the Galaxy still presents us with
numerous open questions. The standard lore goes as follows: if the
sources of galactic cosmic rays (possibly but not necessarily
supernova remnants) inject a spectrum $Q(E)\propto E^{-\gamma}$ with
$\gamma\approx 2.1-2.4$, then diffusion of these cosmic rays in the
magnetic field of the Galaxy leads to an equilibrium spectrum which is
$n(E)\propto E^{-\gamma-\delta}$, where the diffusion coefficient is
taken in the form $D(E)\propto E^\delta$. For a Kolmogorov spectrum of
magnetic fluctuations $\delta=1/3$, while for a Kraichnan spectrum
$\delta=1/2$. Of course these statements apply at energies lower than  
the maximum energy of the accelerated particles,
which for protons is expected to be $\sim 10^{15}-10^{16}$
eV. However, if the maximum energy of the accelerated particles were
much larger, in principle the same conclusions would extend up to the
energy for which the Larmor radius equals the coherence scale of the
field, which is typically taken to be $\sim 100$ pc. This corresponds
to energy $\sim (1-3)\times 10^{17}$ eV for a magnetic field $1-3 \mu
G$. The simulations illustrated in this paper can be performed for
proton energies $E>10^{15}$ eV (in a few cases $E>10^{14}$ eV),
therefore for at least two decades in 
energy we should be able to test the standard lore sketched above. We
confirm that this is the case by considering a toy model with only a
turbulent field with given power spectrum, in which case we are in
perfect agreement with the expectations. The problems arise as soon as
any complication is added to this simple scenario. We illustrate our
points by considering other three toy models, each having a specific
feature which is supposed to be resemblant of a corresponding feature
expected to be present in the actual Galactic magnetic field. In
particular we consider a benchmark situation in which the Galactic
field is taken to have a perfectly azimuthal geometry, so that the
magnetic field lines are closed loops. We showed how in such a
geometry the role of perpendicular diffusion in the escape of
particles from the toy Galaxy is crucial and leads to escape times
which are too long to be reconciled with the observed confinement
times at much lower energies. This conclusion should remain valid in
the case in which the magnetic field lines follow the spiral arms
rather than being closed, since the arms are in any case much larger
in length than the size of the halo. 

An important piece of information should be added: the escape times that
we plotted throughout the paper are all meant to be the average of the
log of the escape times. The spread around these mean values are very
large, covering about one order of magnitude. 
Such spreads
do not reflect limitations in the statistics of the propagated
particles: they are rather stable if the number of particles is
increased. The fluctuations are due to the several possible histories
that may characterize the propagation of cosmic rays in the
Galaxy. On the other hand, the mean values used to infer our conclusions
are very stable.

Another important ingredient of the magnetic field configuration with
closed magnetic field lines consists of the drift motions induced on
the particles by the gradients in the direction of the local large
scale field. The effect of drift is especially evident for high
energies and low levels of turbulence. Similar drifts are introduced
by gradients in the $z$ and $\rho$ directions. 

The most important conclusion that we could achieve is that the
dominant role of the perpendicular diffusion in a geometry with a
prominent azimuthal magnetic field makes the expectation of the common
lore hard to realize. The energy dependence of the perpendicular
diffusion coefficient is not the same as that of the parallel
diffusion coefficient: more specifically in the lower energy regime it
scales as $D_\perp\propto E^{0.5-0.6}$, rather than $E^{1/3}$ as would be
expected for a Kolmogorov spectrum. 
Unfortunately we are not able to follow this behaviour down to
energies below $10^{15}$ eV. In any case, at $\sim 10^{15}$ eV, the
escape times that we {\it measured} in the simulation are always too
large to be extrapolated down to the few million years inferred from
the abundance of light elements in the GeV energy region, even
admitting that a flattening to a behaviour $\propto E^{-1/3}$ of the
escape times could be achieved below $\sim 10^{15}$ eV. 

It is interesting to speculate about possible physical effects that
might cause the escape time to be reduced. From the discussion above,
it is clear that reducing the level of turbulence (namely the value of
$\delta B/B$) does not help, since this would cause the perpendicular
diffusion to decrease, thereby increasing the escape times even
more. Making the halo have a smaller scale height does help, but it
appears rather unrealistic to reduce this scale below $0.5$ kpc
(observations of the radio background from synchrotron emission of
relativistic electrons hint to a typical scale height of a few kpc
\cite{radio}). One possibility that we will discuss more
quantitatively in Paper II is that of a galactic wind, possibly
injected by cosmic rays themselves: in this case, in addition to the
diffusive motion, particles would have a systematic drift velocity
pushing them away from the disc of the Galaxy. 
If the typical wind velocity is $u_W\sim 10^7\rm cm~s^{-1}$
\cite{wind}, then the typical escape time due to advection is of order
5 million years, independent of momentum. It is important to notice
that for the diffusion coefficients used in the literature (in the
common lore usually one does not distinguish between parallel and
perpendicular) this is roughly the escape time scale for cosmic rays
in the GeV energy region, therefore the effect of the wind is usually
relevant only for low energy particles (at higher energies the escape
is dominated by diffusion). In the scenarios that we find, escape is
due to perpendicular diffusion, and takes place on much longer time
scales as we have seen, therefore the effect of the wind can be that
of producing a roughly energy independent escape time of the order of
$\sim 5$ million years. Unfortunately this does not appear
to be the correct, or at least the complete, picture either. In fact
the escape time is observed to be a function of energy $\tau\propto
E^{-0.6}$, as shown by the energy dependence of the secondary to
primary ratio (e.g. \cite{simpson}), although these measurements have
so far been carried out only up to energies of the order of
$10^4-10^5$ MeV/nucleon.

\section*{Acknowledgments}
T.S. and D.D.M. wish to acknowledge useful discussions with
Randy Jokipii, Joe Giacalone and Bill Matthaeus. The work of D.D.M. and
T.S. is funded in part by NASA APT grant NNG04GK86G. The work of P.B. is
partially funded through grant PRIN-2004. 

\appendix
 
\section{Calculation of the diffusion coefficients in the azimuthal
field}\label{appendix}

We consider a regular magnetic field with constant magnitude and
azimuthal direction as in \S \ref{sec:toy2}. We inject particles at
$\rho_0=8.5\kpc$ or $\rho_0=85\kpc$ and we propagate them for $1\Mpc$
recording their trajectories. In this case we cannot use
Eq.~\ref{eq:de} to calculate the diffusion coefficients since the
average values of the displacements we are considering are no longer 0
due to the drifts.

We proceed as follows: we build histograms of the particle positions at
fixed times during the propagation and then we fit these histograms with
gaussian distributions. The fitted value of the variance allows us to
estimate the diffusion coefficient while the mean value of the
distribution is related to the drift velocity.

The three directions we used to calculate the diffusion coefficients are:
$z$, $\rho$ and $\phi$. The first two correspond to the two
perpendicular coefficients and the latter to the parallel one. For $z$
we simply histogram the $z$ coordinate of the particles. For $\rho$ we
histogram the $\rho=\sqrt{x^2+y^2}$ and then we divide each bin in the
histogram by $\rho$ to take into account the volume element (in
cylindrical coordinates). For $\phi$ we histogram $\phi\times\rho_0$,
where $\rho_0$ is the distance at which the particles were injected.

\begin{figure}
  \centering
  \includegraphics[width=0.9\textwidth]{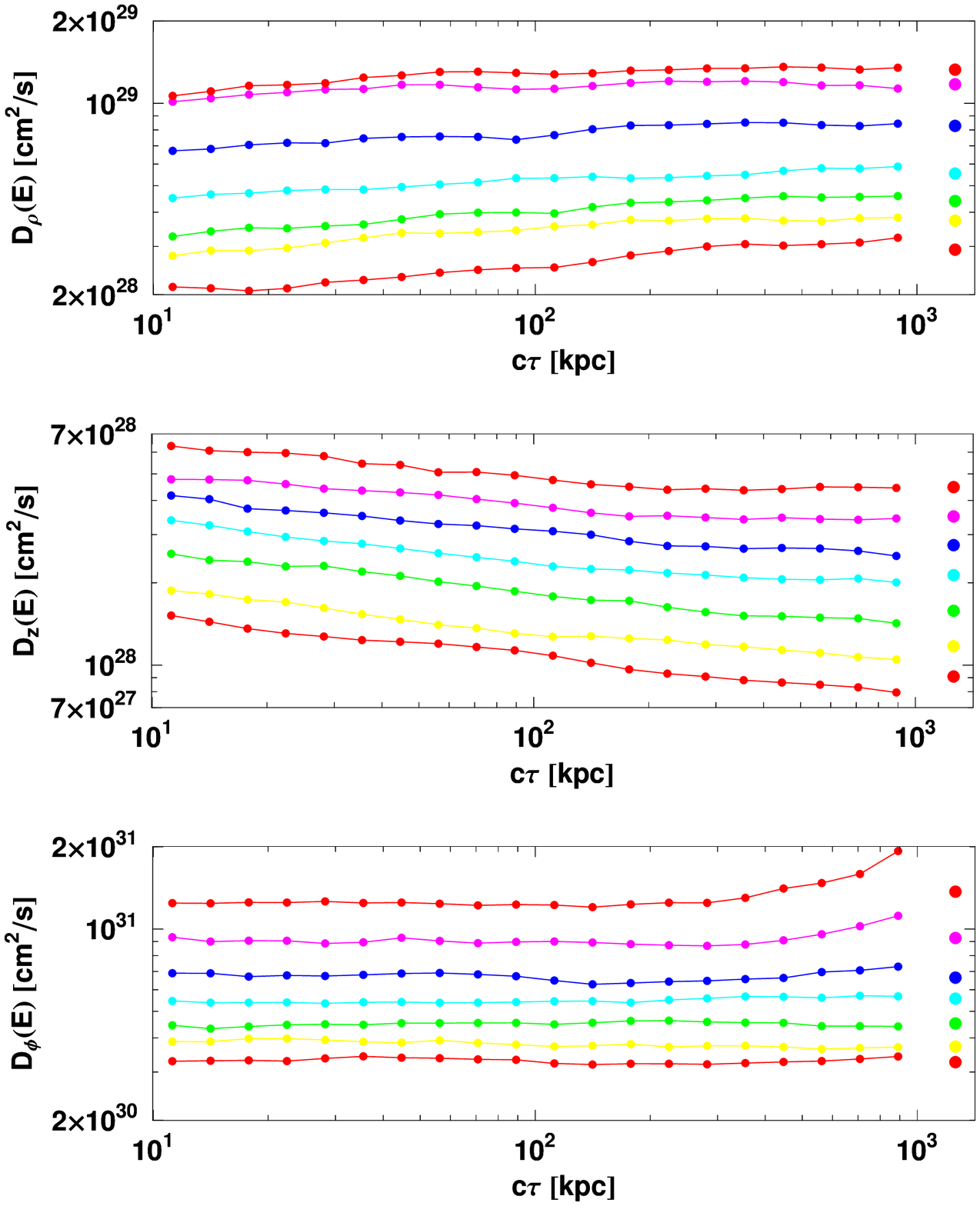}
  \caption{Instantaneous diffusion coefficients as a function of
  propagation time for injection at $8.5\kpc$ in a field composed of
  large scale azimuthal field and an isotropic turbulent field with
  $\delta B/B_0=0.5$.  The three panels, from top to bottom, show the
  three diffusion coefficients along $\rho$, $z$ and $\phi$
  respectively.}\label{fig:diffaz1}
\end{figure}

In Fig.~\ref{fig:diffaz1} we plot the three diffusion coefficients as a
function of propagation time for injection at $8.5\kpc$ and $\delta
B/B=0.5$. The top panel represents the diffusion coefficient in the
$\rho$ direction, the middle panel the one in the $z$ direction and the
lower panel the parallel diffusion coefficient. The differently colored
lines represent, from bottom to top, different energies from
$10^{15}\eV$ to $10^{16.2}\eV$ with a logarithmic step of $0.2$. The
points to the far right of the plots are the average of the last 10
corresponding points and represent our estimate of the diffusion
coefficient.

Concerning the parallel diffusion coefficient we can see that the curves
are flat and that the diffusion regime is achieved. The only unexpected
feature is in the two highest energy curves, corresponding to $10^{16}$
and $10^{16.2}\eV$ that show a steepening around $c\tau\simeq1\Mpc$.
This steepening is simply due to the fact that at high energies and
large propagation times some of the particles have enough time to
complete half a circle around the ``galaxy'' and since to measure the
parallel displacement we are using $\phi\times\rho_0$, particles with
$|\phi|>\pi$ end up in the wrong place in the histogram and distort the
distribution. This is not however a physical effect and it is just a
glitch of the method used to estimate the parallel displacement and we
can just trow away the last few points and do the average with the
remaining ones.

The diffusion coefficient along $z$ shows a tiny sub-diffusion at low
energies, $D_z(E,\tau)\propto \tau^{-0.15}$, that disappears by
increasing the particle energy. It is interesting to note that in the plots of
Fig.~\ref{fig:idc}, that were obtained for similar parameters, but with
the large scale field constant and along the $z$ direction, the
diffusion regime was obtained already with $c\tau\sim10\kpc$, with
slightly larger times necessary for higher energies. In the present
situation the results show that the opposite is occurring: at high energy the
particles reach the diffusion regime, whereas at low energy they may
not, at given time. 
In this case, since the instantaneous diffusion coefficient shows
sub-diffusion, it is not completely correct to define a diffusion
coefficient using the average of the last few points. We do it anyway
averaging the points with propagation times between $100\kpc$ and
$1\Mpc$ that represent the range of propagation times obtained in
Fig.~\ref{fig:Baz_times} for energies between $10^{15}\eV$ and
$10^{17}\eV$. In this way we obtain at least a rough estimate of the
diffusion coefficient affecting the particle propagation in our specific
case.

For the diffusion coefficient in the $\rho$ direction we have a
situation similar to the $z$ one, but with super-diffusion instead of
sub-diffusion. In this case the effect is even smaller with:
$D_\rho(E,\tau)\propto \tau^{0.1}$. Again increasing the energy the
anomalous diffusion disappears.

Increasing the injection distance to $85\kpc$ the anomalous diffusion
is reduced, but it is still slightly present. On the other hand
increasing the turbulence level to $\delta B/B_0 = 1$ or $2$ reduces it
much more than increasing the distance.

\begin{figure}
  \centering
  \includegraphics[width=0.9\textwidth]{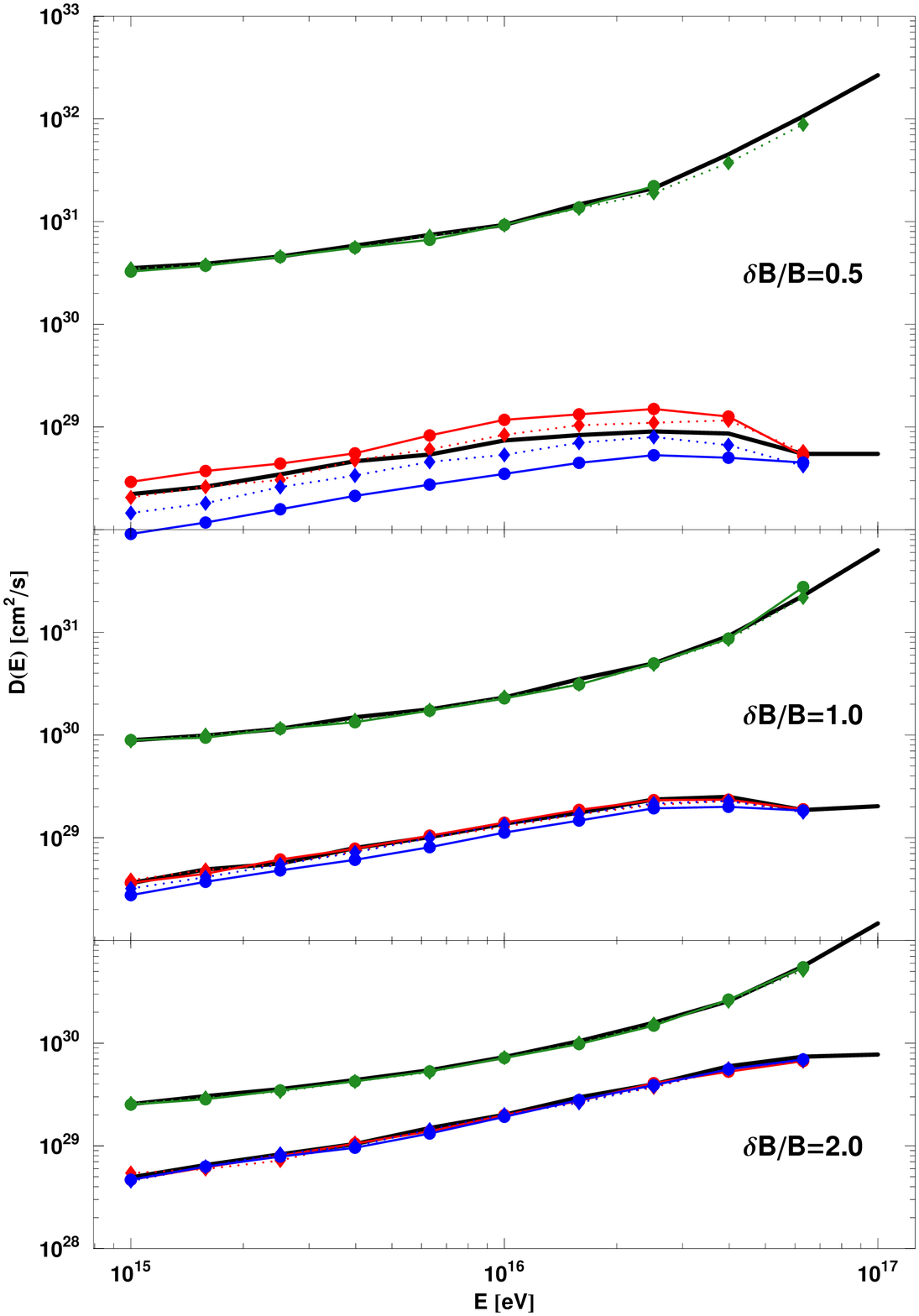}
  \caption{Diffusion coefficients as a function of energy. Green lines:
  parallel. Red lines: along $\rho$. Blue lines: along $z$. Black lines:
  parallel and perpendicular diffusion coefficients from
  Fig.~\ref{fig:d_e}. Solid lines: injection at $8.5\kpc$. Dotted lines:
  injection at $85\kpc$. The three levels of turbulence used are
  indicated in the panels.}\label{fig:daz_e}
\end{figure}

This is clear in the plots of Fig.~\ref{fig:daz_e} where we report the
diffusion coefficients as a function of energy for the three levels of
turbulence and the two injection distances. The black thick lines are
the results for the case of constant large scale field directed along
$z$ (the curves of Fig.~\ref{fig:d_e}). The green lines are the
diffusion coefficients in the $\phi$ direction, the parallel ones. The
red and blue lines are the diffusion coefficients in the $\rho$ and $z$
direction respectively. The solid lines are for injection at $8.5\kpc$
and the dotted ones for injection at $85\kpc$.

The above caveat about anomalous diffusion notwithstanding, the plots in
Fig.~\ref{fig:daz_e} seem to explain the results of
Fig.~\ref{fig:Baz_times}. For the case of injection at $8.5\kpc$ and
$\delta B/B_0=0.5$ we found in \S \ref{sec:toy2} that the obtained escape
time was bigger that the one we expected from the perpendicular
diffusion coefficient. This is consistent with the results presented in
Fig.~\ref{fig:daz_e} where it is shown that, in this case, the diffusion
coefficient in the $z$ direction is reduced and this obviously produces
an increase in the escape time. Increasing the injection distance
the $z$ diffusion coefficient is closer to the ``unmodified'' one (see
dotted blue line in the top panel of Fig.~\ref{fig:daz_e}) and the times
of escape are almost on top of the expectations (see light blue
triangles and top black line in Fig.~\ref{fig:Baz_times}).

Increasing the level of turbulence, the effect of the curvature of the
field lines is reduced and both the $\rho$ and $z$ diffusion
coefficients rapidly converge towards the normal one (middle and bottom
panels in Fig.~\ref{fig:daz_e}).

\end{document}